\documentclass[aps,prx,reprint,superscriptaddress,amsmath,amssymb,longbibliography,floatfix]{revtex4-2}
\usepackage{amsmath,amssymb,amsthm}
\usepackage{mathtools}
\usepackage{graphicx,xcolor}
\usepackage{needspace}
\usepackage{tikz}
\usepackage{hyperref}
\newcommand{\Tr}{\operatorname{Tr}}
\newcommand{\norm}[1]{\left\lVert#1\right\rVert}
\newcommand{\ket}[1]{\lvert#1\rangle}
\newcommand{\bra}[1]{\langle#1\rvert}
\newcommand{\proj}[1]{\ket{#1}\!\bra{#1}}
\newcommand{\LC}{\mathrm{LC}}
\newcommand{\diam}{\operatorname{diam}}
\newcommand{\supp}{\operatorname{supp}}
\newcommand{\one}{\mathbf 1}
\newtheorem{theorem}{Theorem}
\newtheorem{proposition}[theorem]{Proposition}

\usepackage[framemethod=tikz]{mdframed}
\surroundwithmdframed[
 linewidth=0.4pt,linecolor=black,backgroundcolor=white,
 leftmargin=-4pt,rightmargin=-4pt,
 innerleftmargin=3.6pt,innerrightmargin=3.6pt,
 innertopmargin=3pt,innerbottommargin=3pt,
 skipabove=6pt,skipbelow=6pt,nobreak=true
]{theorem}

\hypersetup{pdftitle={Local-Channel Stability of Classical Z2 Lattice Gauge Models},pdfauthor={Guanxi Xun, Weiguang Cao, Haruki Watanabe},pdfsubject={Fixed spatial dimension and spins on cells of general dimension; stability under weak classical and quantum perturbations}}
\begin{document}
\title{Local-Channel Stability of Classical \texorpdfstring{$\mathbb Z_2$}{Z2} Lattice Gauge Models}
\author{Guanxi Xun}
\affiliation{Department of Physics, Hong Kong University of Science and Technology, Clear Water Bay, Hong Kong, China}

\author{Weiguang Cao}
\affiliation{Department of Physics, Hong Kong University of Science and Technology, Clear Water Bay, Hong Kong, China}
\affiliation{Center for Theoretical Condensed Matter Physics, Hong Kong University of Science and Technology, Clear Water Bay, Hong Kong, China}

\author{Haruki Watanabe}
\email{hwatanabe@ust.hk}
\affiliation{Department of Physics, Hong Kong University of Science and Technology, Clear Water Bay, Hong Kong, China}
\affiliation{Center for Theoretical Condensed Matter Physics, Hong Kong University of Science and Technology, Clear Water Bay, Hong Kong, China}
\affiliation{Institute for Advanced Study, Hong Kong University of Science and Technology, Clear Water Bay, Hong Kong, China}
\date{\today}
\begin{abstract}
We establish sufficient conditions for two-way local quantum-channel stability in classical $\mathbb Z_2$ lattice gauge models with spins on $m$-dimensional cells of a $d$-dimensional lattice. Each interaction multiplies the spins on the boundary of an $(m+1)$-cell. This family, defined on the full spin Hilbert space without a Gauss-law constraint, includes the Ising model at $m=0$ and the magnetic sector of the toric code at $m=1$. The low-temperature proof combines two ingredients: locally indistinguishable flat sectors and extended magnetic defects of dimension $d-m-1$. For $1\le m\le d-2$, sufficiently weak local classical perturbations at sufficiently low positive temperature preserve equivalence to the canonical flat ensemble. No internal or gauge symmetry is required of the perturbation. For weak noncommuting perturbations, we prove the same conclusion in a smaller sufficient window by reducing the diagonal probabilities to a weak classical interaction and recovering the coherences with a local channel. The conversion channels use product ancillary inputs, have polylogarithmic causal radius, and achieve any prescribed inverse-power whole-system trace-norm error. Sufficiently high-temperature states are instead two-way equivalent to a product state. Independent finite-dimensional onsite systems can be included with weak local couplings. An optimized covariance between two distant neighborhoods of parallel $m$-tori is nonincreasing under one-way local conversion and invariant under two-way conversion. Its values $\mathcal F_m=1$ and $0$ distinguish the proved low- and high-temperature domains without a protecting symmetry. These sufficient domains do not determine the full phase diagram.
\end{abstract}
\maketitle
\section{Introduction}
\label{sec:introduction}

A classical Gibbs state can carry information that two distant, extended regions read in common, even when every bounded region is insensitive to that information. The familiar example is the plaquette $\mathbb Z_2$ gauge model: two parallel loops winding around a periodic system can read the same random sign. At low temperature, a reader can first correct small flux defects and then recover that sign~\cite{Watanabe2026}. This information is classical---the unperturbed state is diagonal and separable---but creating it from independent local inputs requires communication across the system.

We study the stability of this information through local quantum channels. Two state families are equivalent if channels convert them in both directions with vanishing whole-system trace-norm error and sublinear causal radius. A channel may discard information, so one-way preparation or decoding alone does not establish equivalence. Because ancillary inputs are products over local sites, they cannot supply a common random winding sign externally. These conventions place the present problem within the local-operation approach to mixed-state phases~\cite{CoserPerezGarcia2019,SangZouHsieh2024}.

Spins on lattice cells place the Ising and plaquette models in one family~\cite{Wegner1971} [Fig.~\ref{fig:dimension-range}(a)]. In a $d$-dimensional cubic lattice, put a qubit on each $m$-dimensional cell and multiply its $Z$ operators around each elementary $(m+1)$-cell. At $m=0$, the interaction is $Z_iZ_j$ on a link, giving the ordinary ferromagnetic Ising model. At $m=1$, it is the plaquette interaction of the toric code's magnetic sector~\cite{Kitaev2003}, with link qubits and no electric star term. Higher $m$ replaces loops and membranes by higher-dimensional cell sets without changing the basic mechanism.

This family separates two requirements that coincide in the 3D link model. First, $m\ge1$ makes the flat-sector information nonlocal: the uniform flat ensembles in different winding sectors have the same marginal on a bounded contractible region. Second, $d-m-1\ge1$ makes magnetic defects extended. Their energy grows with their size, and low temperature suppresses components large enough to spoil local correction. Both requirements are needed for the stability result under arbitrary weak local perturbations. Ising domain walls are extended, but a local magnetic field can distinguish and select their ordered sectors. Conversely, at $m=d-1$ the magnetic defects are points. The range proved here is therefore $1\le m\le d-2$, which requires spatial dimension $d\ge3$.

For the 3D link model, classical memory under gauge-breaking fields has been studied numerically~\cite{PoulinMelkoHastings2019} and through slow-mixing bounds~\cite{StahlEtAl2026}. Gauge averaging provides a route to stability under general weak classical perturbations~\cite{Thorngren2026}. Classical topological information and finite-temperature topology have also been studied through entropy diagnostics~\cite{CastelnovoChamon2007,CastelnovoChamon2008}. Related sampling results distinguish rapid equilibration within a logical sector from slow relaxation of the logical information~\cite{BergamaschiGheissariLiu2025}.

Our question is stronger than the persistence of a chosen observable or a long relaxation time: can the full perturbed Gibbs state be obtained from, and converted back to, the unperturbed reference by local operations? A complementary approach uses finite Markov length, meaning exponential decay of conditional mutual information across a widening buffer. Uniform control during a local evolution enables quasi-local recovery~\cite{SangHsieh2025}, while axiomatic approaches characterize topological mixed states through recoverability and correlation conditions~\cite{YangShiLee2026}. Local circuits connecting Gibbs states along a Hamiltonian path have also been constructed under stable-clustering assumptions~\cite{MaKhemaniSang2026}. Here we verify explicit sufficient domains by controlling defects and constructing both conversion directions.

Building on the 3D invariant and decoding diagnostics of Ref.~\cite{Watanabe2026}, we construct both channel directions in sufficient domains for the stated $(d,m)$ range, with controlled whole-system trace-norm error, including coherence recovery under noncommuting perturbations, high-temperature preparation, and weak coupling to physical ancillary systems. The classical construction first averages over local gauge transformations, then removes or samples dilute connected defect components, and finally reverses the gauge average on the specified state. For noncommuting perturbations, a convergent imaginary-time expansion reduces the diagonal probabilities to a weak classical interaction. A separate, deterministic local channel restores the coherences, using spectral flow of auxiliary Hamiltonians with a product reference gap. At high temperature a related auxiliary construction prepares the Gibbs state directly from local product inputs.

We extend the optimized winding-correlation invariant of Ref.~\cite{Watanabe2026} to an $m$-cell quantity $\mathcal F_m$ that makes the resulting phase distinction explicit. The two readers occupy distant neighborhoods of parallel primitive $m$-tori. They can use arbitrary regional observables, including local decoding, rather than only bare Wilson surfaces. One-way local conversion cannot increase $\mathcal F_m$, so two-way conversion preserves it. Its values are one in the established low-temperature domains and zero in the high-temperature domain. These sufficient domains do not specify a critical temperature or exhaust either phase.

The companion Letter~\cite{WatanabeThermodynamics2026} studies free-energy analyticity for the same cell models, with a fixed finite spatial period. The two works share the model and geometric preliminaries, local fillings, gauge averaging, weighted classical expansions, and the imaginary-time reduction of diagonal weights. Here these ingredients support two-way conversion of full states and the phase distinction measured by $\mathcal F_m$, without spatial periodicity. This requires additional conditional recovery, sampling, and coherence-recovery constructions with global error and causal-radius bounds. We identify the shared steps and give complete proofs in the Supplemental Material (SM)~\cite{SupplementalMaterial}, while retaining the conversion mechanisms and their physical interpretation in the main text.

Section~\ref{sec:results} states the model and conversion theorems. Sections~\ref{sec:classical} and~\ref{sec:quantum} explain classical transport and quantum coherence recovery. Section~\ref{sec:high-ancilla} treats high temperature and physical ancillary systems, and Sec.~\ref{sec:invariant} defines the winding invariant. The SM supplies the detailed estimates and circuit constructions.

\section{Model and local-channel stability}
\label{sec:results}
We first define the spin models, reference states, and allowed local channels. We then state the low- and high-temperature results.

\subsection{Spin models on \texorpdfstring{$m$}{m}-cells}

A $j$-cell is a $j$-dimensional elementary cell of the cubic lattice: a vertex for $j=0$, a link for $j=1$, a plaquette for $j=2$, and a cube for $j=3$. Take a periodic hypercubic lattice of integer spatial dimension $d\ge1$ and side length $L$, choose $0\le m\le d-1$, and let $\Lambda_L^{(j)}$ denote its $j$-cells. Define
\begin{equation}
 N\coloneqq L^d,\qquad n_j\coloneqq\binom dj N,\qquad n\coloneqq n_m.
 \label{eq:counts}
\end{equation}
There are $N$ unit cells and $n$ physical qubits. The parameters $d,m$, and all local interaction bounds are fixed before $L\to\infty$. A physical qubit on each $m$-cell $\ell$ has Pauli operators $Z_\ell,X_\ell$, where $X_\ell$ flips the $Z_\ell$ eigenvalue. For an $(m+1)$-cell $p$, let $\partial p$ be its $2(m+1)$ boundary $m$-cells and set
\begin{align}
 B_p&\coloneqq\prod_{\ell\in\partial p}Z_\ell,\\
 H_{d,m}&\coloneqq-J\sum_{p\in\Lambda_L^{(m+1)}}B_p,
 \qquad J>0.\label{eq:model}
\end{align}
For fixed $d,m$, write $H_0\coloneqq H_{d,m}$. The family is classical because all $B_p$ commute and are diagonal in the product $Z$ basis. Orientations are unnecessary for two-state spins. We set the lattice spacing and Boltzmann's constant to one and define $\beta\coloneqq1/T$. Both $J$ and the perturbation strength $g$ below have units of energy.

\begin{figure*}[t]
\centering
\includegraphics[width=410.30325pt,height=126.38626pt]{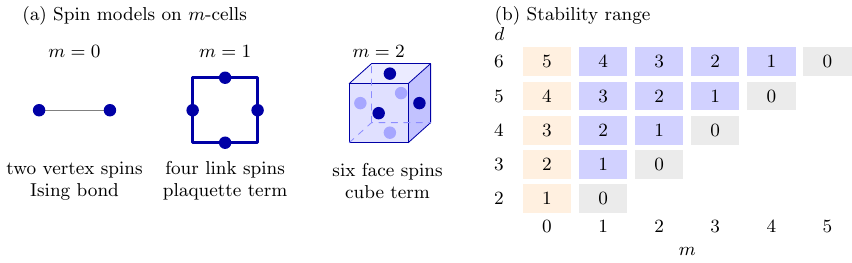}
\caption{Family of spin models defined on $m$-cells and its stability range.
(a) Each interaction multiplies the spins on the boundary of one cell.
Spins occupy vertices, links, or faces for $m=0,1,2$, respectively.
Dots mark spin centers.
(b) Entries give the magnetic-defect dimension $d-m-1$.
Blue cells satisfy $1\le m\le d-2$, the range of our low-temperature stability results.
Orange marks vertex-spin Ising models; gray marks point-defect models.
These colors label model families, not thermal phases.
The model schematic in (a) is shared with Ref.~\cite{WatanabeThermodynamics2026}.}
\label{fig:dimension-range}
\end{figure*}

At $m=0$, $p$ is a link $\langle ij\rangle$ whose boundary is its two endpoints, and
\begin{equation}
 H_{d,0}=-J\sum_{\langle ij\rangle}Z_iZ_j.
 \label{eq:ising}
\end{equation}
Here $i,j$ label neighboring lattice vertices. No rescaling or additive constant is needed. At $m=1$, the same rule gives the four-link plaquette interaction
\begin{equation}
 H_{d,1}=-J\sum_p\prod_{\ell\in\partial p}Z_\ell.
 \label{eq:link-model}
\end{equation}
Equation~\eqref{eq:link-model} is the magnetic, or $Z$-plaquette, sector of the $d$-dimensional toric code, without the electric star term. Figure~\ref{fig:dimension-range}(a) illustrates the first three choices of $m$.

Throughout, ``lattice gauge model'' is conventional terminology for this spin model on the full tensor-product Hilbert space. We impose no Gauss-law constraint, do not identify configurations related by gauge transformations, and do not fix a winding sector. A weak electric star term is allowed as a perturbation, but no claim about finite-temperature quantum memory of the full toric code follows from the results below.

The physical spin signs are $\sigma\coloneqq(\sigma_{\ell})_{\ell}$, with $\sigma_{\ell}\in\{+1,-1\}$. For parity operations and product-basis labels we also use binary variables $a_{\ell}\in\{0,1\}$ defined by $\sigma_{\ell}\coloneqq(-1)^{a_{\ell}}$, and write $a\coloneqq(a_{\ell})_{\ell}$. Thus $\ket a$ and $\ket\sigma$ denote the same basis vector. The binary notation is convenient for recording spin flips as addition modulo two. The interaction value is $B_p(a)\coloneqq\prod_{\ell\in\partial p}(-1)^{a_\ell}$.

A configuration is \emph{flat} when $B_p(a)=1$ for every $p$. Let $\mathcal A_{\rm flat,L}$ be the set of all flat configurations, and define the reference states
\begin{align}
 \rho_{\rm flat,L}&\coloneqq\frac1{|\mathcal A_{\rm flat,L}|}
       \sum_{a\in\mathcal A_{\rm flat,L}}\proj a,\\
 \rho_{\infty,L}&\coloneqq2^{-n}I.
 \label{eq:references}
\end{align}
The first state is the canonical $T\downarrow0$ limit of $H_0$, with all winding sectors included. The second is an onsite product state.

\subsection{Local gauge symmetry and wrapping transformations}

For $m\ge1$, signs $\eta_u\in\{+1,-1\}$ on the $(m-1)$-cells parametrize local gauge transformations,
\begin{equation}
 (\eta\sigma)_\ell\coloneqq\sigma_\ell
                \prod_{u\in\partial\ell}\eta_u.
 \label{eq:gauge-action}
\end{equation}
Here $\partial\ell$ consists of the $2m$ boundary $(m-1)$-cells of $\ell$. Every $B_p$, and hence $H_0$, is invariant. These transformations are symmetries of the unperturbed spin Hamiltonian. Configurations related by them remain distinct states in the full trace. The proof uses this invariance to average a perturbing Boltzmann weight. It does not impose the same invariance on the perturbation.

To describe wrapping transformations and defect fillings, we use cell sets with addition modulo two. For a set of $j$-cells with $j\ge1$, its boundary consists of the $(j-1)$-faces occurring an odd number of times. A set of vertices has empty boundary. A set is closed when its boundary is empty. For dual cells, the boundary is taken in the dual lattice. For two cell sets $A,B$, their symmetric difference $A\triangle B\coloneqq(A\setminus B)\cup(B\setminus A)$ contains the cells belonging to exactly one set. Adding cell sets means taking this symmetric difference, so repeated cells cancel in pairs. In particular, the boundary of a boundary is empty.

To distinguish this local symmetry from the symmetry that characterizes unperturbed order, let $\Sigma$ be a closed set of dual $(d-m)$-cells and $\ell^\ast$ the dual cell crossing $\ell$. Define
\begin{equation}
 U_\Sigma\coloneqq\prod_{\ell:\,\ell^\ast\in\Sigma}X_\ell.
 \label{eq:form-symmetry}
\end{equation}
This commutes with $H_0$ because each interaction meets an even number of flipped spins. For a closed physical $m$-dimensional cell set $C$, the Wilson product is
\begin{equation}
 W_C\coloneqq\prod_{\ell\in C}Z_\ell.
 \label{eq:Wilson}
\end{equation}
Conjugation by $U_\Sigma$ reverses its sign precisely when $C$ and $\Sigma$ intersect an odd number of times. For $m\ge1$, a bounding $\Sigma$ gives a product of the local gauge transformations in Eq.~\eqref{eq:gauge-action}. Wrapping $\Sigma$ generate the $m$-form symmetry modulo those local transformations.

For $0\le m\le d-2$, the sufficiently low-temperature phase of $H_0$ admits the interpretation of spontaneous breaking of this wrapping $m$-form symmetry~\cite{Wegner1971,GaiottoEtAl2015}. For $m\ge1$, large contractible Wilson products obey a generalized perimeter law: their logarithmic decay scales with their $m$-dimensional support rather than the enclosed $(m+1)$-dimensional volume. Removing that local support-dependent decay is the usual higher-form order criterion. At $m=0$ this reduces to Ising order. This symmetry breaking is defined in the thermodynamic limit. The local gauge transformations of $H_0$ remain unbroken~\cite{Elitzur1975}. The finite-volume canonical state remains symmetric, with zero mean for charged winding observables. The channel criterion below applies even when a perturbation explicitly breaks the wrapping symmetry.

\subsection{Local perturbations}

The perturbed Hamiltonian and Gibbs state are
\begin{align}
 H_g&\coloneqq H_0+gV,\\
 V&\coloneqq\sum_Xv_X,\\
 Z_L(\beta,g)&\coloneqq\Tr e^{-\beta H_g},\label{eq:partition}\\
 \rho_{\beta,g,L}&\coloneqq Z_L(\beta,g)^{-1}e^{-\beta H_g}.
 \label{eq:perturbed}
\end{align}
The trace in Eq.~\eqref{eq:partition} runs over all $2^n$ states. We use $\norm{O}$ for operator norm, $\norm{\rho-\sigma}_1$ for trace norm, and $\norm{f}_\infty$ for a function's supremum norm. Each dimensionless $v_X$ is Hermitian with $\norm{v_X}\le1$ and acts on a nonempty assigned set $X$ of at most $b$ physical $m$-cells and diameter at most $R$. At most $n_{\mathrm{inc}}$ assigned supports contain any given physical cell. The local operators are independent of $\beta,g$, and all locality bounds are independent of $L$. No spatial periodicity is required for the channel results. Unnamed estimate constants can change from line to line and depend on the fixed local data.

We call the perturbation classical when the total $V$ is diagonal in the product $Z$ basis. In that case, each local term can be replaced by its diagonal part without changing $V$ or the locality bounds. In the general quantum case, the terms may fail to commute with one another or with $H_0$. We require no internal symmetry of $V$, no invariance under the local transformations in Eq.~\eqref{eq:gauge-action}, and no invariance under the wrapping $U_\Sigma$. For example, $V=-\sum_\ell Z_\ell$ explicitly breaks the wrapping transformations.

\subsection{Allowed channels and accuracy}

A physical local channel is a circuit of geometrically supported completely positive trace-preserving maps. Work registers start in an input-independent product state over local sites. Classical messages and work registers count toward the causal radius, and no postselection or external shared randomness is allowed. The radius $R_{\Phi_L}$ bounds the expansion of supports in a local dilation of the entire circuit. Block gates may perform arbitrary operations inside their support. We make no claim of efficient elementary-gate decompositions.

For state families $\boldsymbol\rho\coloneqq\{\rho_L\}_L$ and $\boldsymbol\sigma\coloneqq\{\sigma_L\}_L$, write $\boldsymbol\rho\longrightarrow_{\LC}\boldsymbol\sigma$ if channels satisfy
\begin{equation}
 R_{\Phi_L}=o(L),\qquad
 \norm{\Phi_L(\rho_L)-\sigma_L}_1\longrightarrow0,
 \label{eq:lc}
\end{equation}
and write $\boldsymbol\rho\sim_{\LC}\boldsymbol\sigma$ when both directions exist. The channels convert the specified state families in opposite directions. Their composition need not be the identity on arbitrary inputs. Whole-system trace error in Eq.~\eqref{eq:lc} is stronger than agreement of every fixed local marginal.

This endpoint equivalence does not require each local operation along a prescribed evolution to be locally reversible, a stronger condition studied in Ref.~\cite{SangEtAl2025LocalReversibility}. A sufficient recovery criterion controls the conditional mutual information $I(A:C\mid B)\coloneqq S(AB)+S(BC)-S(B)-S(ABC)$, where $B$ is a buffer separating $A$ from the rest $C$ and $S$ is the von Neumann entropy. Its exponential decay with buffer width must hold uniformly through the intermediate operations, including partial circuit layers~\cite{SangHsieh2025}; finite Markov length of the endpoints alone is insufficient. This criterion is distinct from free-energy analyticity: for a full finite-range classical Gibbs state, conditioning on a buffer wider than the interaction range makes $A$ and $C$ independent even at a thermal transition.

% Keep the result heading with the first framed theorem.
\Needspace{24\baselineskip}
\subsection{Main theorem and the range of cell dimensions}

We now state our main local-channel results at low and high temperatures.

\begin{theorem}[Low-temperature local-channel stability]
\label{thm:main}
Fix $d,m$ with $1\le m\le d-2$ and a local interaction class $(b,R,n_{\mathrm{inc}})$. There are positive constants $\tilde B,\tilde a,\tilde C$ such that either of the conditions
\begin{equation}
 \begin{cases}
 \beta J>\tilde B,\quad \beta|g|<\tilde a,
       &V\text{ classical},\\
 \beta J>\tilde B,\quad \beta|g|e^{\tilde C\beta J}<\tilde a,
       &V\text{ Hermitian}
 \end{cases}
 \label{eq:low-domain}
\end{equation}
implies
\begin{equation}
 \boldsymbol\rho_{\beta,g}\sim_{\LC}\boldsymbol\rho_{\rm flat}.
 \label{eq:low-equivalence}
\end{equation}
For every fixed $s>0$, each conversion has whole-system trace-norm error $O(L^{-s})$, polylogarithmic gate diameter, depth, and causal radius, and product work-register inputs.
\end{theorem}
Setting $g=0$ gives $\boldsymbol\rho_{\beta,0}\sim_{\LC}\boldsymbol\rho_{\rm flat}$. Composing the conversions therefore yields $\boldsymbol\rho_{\beta,g}\sim_{\LC}\boldsymbol\rho_{\beta,0}$ at the same temperature, with the same error and locality bounds.
% Put surplus column space below the frame rather than around the heading.
\par\vfill

\begin{theorem}[High-temperature local preparation]
\label{thm:high}
For any fixed $d,m$ with $0\le m\le d-1$ and bounded finite-range interaction class above, there is $\tilde a'>0$ such that
\begin{equation}
 \beta[2(d-m)J+n_{\mathrm{inc}}|g|]<\tilde a'
 \label{eq:high-domain}
\end{equation}
implies
\begin{equation}
 \boldsymbol\rho_{\beta,g}\sim_{\LC}\boldsymbol\rho_\infty.
 \label{eq:high-equivalence}
\end{equation}
Both directions obey the same locality, product-ancilla, and whole-system accuracy guarantees as in Theorem~\ref{thm:main}.
\end{theorem}

The constants depend on $d,m$ and the interaction class. They need not remain uniform as the dimension grows. In Eq.~\eqref{eq:high-domain}, $2(d-m)$ counts the interaction cells meeting each physical $m$-cell. The preparation threshold $\tilde a'$ also contains dimension-dependent geometric estimates for the local channels.

The high-temperature statement applies outside the low-temperature range and reaches sufficiently high temperature at any fixed finite $g$. For fixed nonzero $g$, the low-temperature windows in Eq.~\eqref{eq:low-domain} do not reach $T=0$. Neither construction assumes a gap in the physical Hamiltonian $H_g$. Theorem~\ref{thm:ancilla} below extends both regimes to the full joint state with independent onsite physical ancillary systems and weak local couplings.

The range $1\le m\le d-2$ has two separate geometric origins [Fig.~\ref{fig:dimension-range}(b)]. For $m\ge1$, flat sectors are labeled by winding signs measured on noncontractible $m$-dimensional surfaces, and a bounded contractible region cannot distinguish their uniform ensembles. For $m\le d-2$, the magnetic defects have positive dimension and admit local fillings when their components are small. The next section explains both assertions using only cell sets and parity.

At $m=0$, the flat reference mixes the two constant Ising configurations. Its random sign is locally visible, so a symmetry-breaking field can select a sector. This case can have conventional low-temperature order, but it is outside Theorem~\ref{thm:main}. At $m=d-1$, defects are points. A single point component has no local filling, so the extended-defect argument fails. The unperturbed thermodynamic free energy per unit cell at this endpoint is exactly
\begin{equation}
 f=-T[d\log2+\log\cosh(J/T)],
 \label{eq:point-defect-free-energy}
\end{equation}
which is analytic at every positive temperature. The Ising chain $(d,m)=(1,0)$ and the 2D plaquette model lie at this endpoint, whereas the 3D plaquette model lies in the stable range. For $d=2$, the only choices in this family are $m=0$ (Ising) and $m=1$ (plaquette), so neither falls within Theorem~\ref{thm:main} for low-temperature stability under arbitrary weak local perturbations. Theorem~\ref{thm:high} still applies to both at sufficiently high temperature. These endpoint exclusions concern the stated stability mechanism, not a classification of every possible perturbation at those endpoints.

\begin{figure}[t]
\centering
\includegraphics[width=179.83421pt,height=158.66818pt]{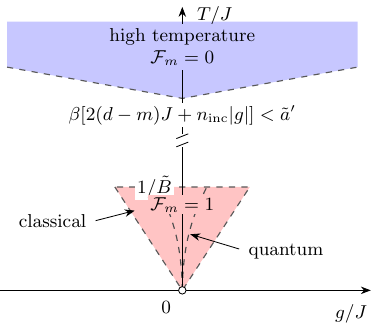}
\caption{Sufficient domains for fixed $1\le m\le d-2$.
The classical low-temperature window is a wedge; the quantum window is narrower.
The high-temperature bound applies to either type of perturbation.
Dashed curves are sufficient bounds, not phase boundaries or numerical
estimates of the constants. The temperature axis is broken and the
intermediate white region is not classified. Positive temperature is required.}
\label{fig:domains}
\end{figure}

Figure~\ref{fig:domains} summarizes the sufficient domains. The exponentially shrinking quantum window reflects the imaginary-time estimates used in the proof, rather than an identified physical phase boundary. We do not extrapolate a 3D duality argument or a result on the whole interval below a critical temperature to arbitrary $(d,m)$.

\section{Defects, flat information, and classical conversion}
\label{sec:classical}
We begin with flux loops in the 3D plaquette model and show explicitly how local spin flips remove them. We then generalize the construction to $1\le m\le d-2$ and organize the two-way conversion under a weak classical perturbation into four steps. Throughout this section the states are diagonal, so all channels can be understood as operations on classical spin configurations.

\subsection{Flux loops and local correction in the 3D model}

Take $(d,m)=(3,1)$, with a spin on each physical link $\ell$ and $B_p$ the product of the four spins around plaquette $p$. A plaquette with $B_p=-1$ costs $2J$ relative to $B_p=+1$. Draw the dual link $p^\ast$ piercing every such plaquette and call the resulting graph $E(a)$. The occupied dual links record the defects, while the physical links carry the spins.

The six plaquette products around a cube multiply to $+1$, because each link spin appears twice. Thus an even number of occupied dual links meets each dual vertex: a flux graph cannot end. A connected component $\gamma$ can be a simple loop or a graph with intersections. It need not be a single nonintersecting polygon. Its excitation energy is $2J|\gamma|$, where $|\gamma|$ counts its dual links. Low temperature therefore suppresses large components through the Boltzmann factor $e^{-2\beta J|\gamma|}$.

\begin{figure}[!tbp]
\centering
\includegraphics[width=214.14554pt,height=74.19717pt]{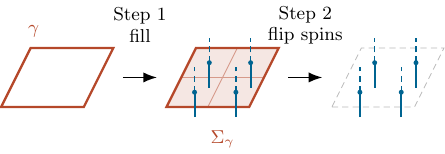}
\caption{Local correction in the 3D plaquette model, $(d,m)=(3,1)$.
The orange loop $\gamma$ consists of dual edges associated with defective
physical plaquettes. Step 1 chooses a surface of dual plaquettes
$\Sigma_\gamma$ with $\partial\Sigma_\gamma=\gamma$.
Step 2 applies $X_\ell$ to each physical link piercing that surface
(blue segments). Each flip toggles the flux on the boundary of its
dual plaquette; interior edges cancel in pairs, leaving exactly
$\gamma$ to be removed. The gray outline records the chosen surface
after correction, not a remaining defect. General small components
need not be planar or simple; their fillings stay inside a local box.}
\label{fp:fig:filling}
\end{figure}

Local spin flips remove a small closed flux component by filling it with dual plaquettes (Fig.~\ref{fp:fig:filling}). First consider a single physical spin flip $X_\ell$. It reverses the four adjacent $B_p$ and hence toggles the four dual links surrounding the dual plaquette $\ell^\ast$. More generally, fill a small closed component $\gamma$ by a set $\Sigma_\gamma$ of dual plaquettes with $\partial\Sigma_\gamma=\gamma$. Apply $X_\ell$ to each physical link piercing this filling. Interior dual links are toggled an even number of times and cancel. Links in $\gamma=\partial\Sigma_\gamma$ are toggled an odd number of times. Exactly the defects in $\gamma$ are removed. The filling need not be planar, and only its boundary changes the defect graph.

To make this correction local, the filling must stay near the component. A component confined to a box that does not wrap around the system has such a filling in that box. The decoder measures the plaquette signs, identifies small connected components, and applies these corrections. Large components are left untreated. The estimate below makes their total probability small.

Once all defects have been removed, the output is flat but still contains the input's nonlocal information: for any fixed defect graph whose components are all small, the correction is a bijection from the configurations with that graph to the full flat set. A uniform input on that set of configurations therefore gives the uniform flat ensemble, without fixing the values of its three independent winding Wilson loops. The reverse channel starts from this flat ensemble and adds randomly sampled small flux components. It need not create a shared winding bit from local randomness.

\subsection{Defect geometry in general cell dimensions}

The same construction applies to spins on $m$-cells. For a configuration $a$, call $p$ defective when $B_p(a)=-1$. Its dual cell $p^\ast$ has dimension
\begin{equation}
 k\coloneqq d-m-1.
 \label{eq:defect-dimension}
\end{equation}
Continue to write $E(a)\coloneqq\{p^\ast:B_p(a)=-1\}$, now a set of defective dual $k$-cells rather than dual links. The lattice Bianchi identity states that the product of the boundary interactions of every $(m+2)$-cell $q$ is one, because each physical $m$-cell occurs twice:
\begin{equation}
 \prod_{p\in\partial q}B_p=I.
 \label{eq:bianchi}
\end{equation}
On the dual lattice, this says $\partial E(a)=\varnothing$. Here $\partial E$ consists of the $(k-1)$-faces incident on an odd number of cells in $E$. In addition, $E(a)$ is the boundary of the dual $(k+1)$-cell set corresponding to the physical cells with $a_{\ell}=1$. Thus physical defect sets obey a global boundary constraint as well as local closure. On a torus, closure alone does not imply this global constraint.

Two occupied dual $k$-cells are adjacent when they share a $(k-1)$-face. A connected component is a maximal set joined by such adjacencies: it includes every occupied cell reachable from any of its cells. Such components may branch or self-intersect. A fixed face sees cells in just one component, so each component is closed. For example, $d=m+2$ gives loops, $d=m+3$ gives sheets, and $d=m+4$ gives 3D defect sets. Every defective interaction costs $2J$:
\begin{equation}
 H_0(a)=-Jn_{m+1}+2J|E(a)|.
 \label{eq:defect-energy}
\end{equation}

The spin configurations with a specified defect set form its defect fiber. Each nonempty fiber has the same number of configurations: if $a$ and $a'$ have the same defects, their bitwise difference is flat, and adding any flat configuration preserves the defects. Hence the unperturbed defect law is
\begin{equation}
 \mathbb P_{\rm phys}(E)\propto
 \one_{\{E\text{ is a boundary}\}}e^{-2\beta J|E|}.
 \label{eq:flux-law}
\end{equation}
The indicator $\one_{\{\cdots\}}$ equals one when the stated condition holds and zero otherwise. The proportionality constant normalizes the sum over all dual $k$-cell sets; only sets that are boundaries receive nonzero weight.

There are positive geometric constants $c_{d,m}$ and $C_{d,m}$, depending only on $d,m$ and independent of $L,\beta,J$, with the following property. A connected closed component $\gamma$ of $j<c_{d,m}L$ cells fits in a nonwrapping box of diameter at most $C_{d,m}j$. Thus $c_{d,m}$ sets a sufficient size cutoff, while $C_{d,m}$ bounds the spatial extent. Since $k\ge1$, it bounds a $(k+1)$-cell set inside that box. One can find such a filling by sweeping the cells successively onto coordinate faces. Closedness cancels the side pieces in each sweep, while the projected copies cancel between successive sweeps. The final projection to a vertex vanishes because $k\ge1$. Fix a deterministic rule that chooses a filling inside this box from $\gamma$ alone. Flipping its corresponding physical $m$-cells removes $\gamma$. Only the filling's diameter, not its number of cells or computational complexity, is needed for the channel bound. SM Sec.~II supplies the precise estimates.

\subsection{An explicit decoder for the unperturbed state}

We show that, at sufficiently low temperature, a local decoder converts the unperturbed thermal state to the uniform flat state, with whole-system trace-norm error vanishing as $L\to\infty$. We first bound the probability of a component too large to correct locally, and then choose a logarithmic size cutoff for the decoder.

\subsubsection{Counting connected defect sets}

The number of dual $k$-cells adjacent to a given one is at most
\begin{equation}
 D_{d,m}\coloneqq2(d-m-1)(2m+3).
 \label{eq:peierls-constants}
\end{equation}
A dual $k$-cell has $2k$ faces of dimension $k-1$. Each face belongs to $2(d-k+1)$ dual $k$-cells, including the original one, giving the bound $2k[2(d-k+1)-1]=D_{d,m}$.

To count connected sets of $j$ cells containing a fixed cell, join their cells by a spanning tree: $j-1$ adjacency links that connect them without a cycle. Fix an ordering to choose one tree and one traversal for each set. Starting at the fixed cell and traversing every tree link twice gives a walk of $2(j-1)$ steps, each with at most $D_{d,m}$ choices. The visited cells recover the original set, so different sets cannot give the same walk. There are therefore at most $D_{d,m}^{2(j-1)}$ such sets. This also bounds the number of possible closed components, since imposing closure can only reduce the count.

\subsubsection{Component probabilities and the boundary constraint}

Deleting a component need not preserve the physical boundary constraint. For example, two parallel winding loops in the 3D model can jointly bound a strip, whereas either loop alone is not a boundary. To allow deletion, temporarily sum over all closed dual $k$-cell sets, including those that are not boundaries, with the same Boltzmann weights:
\begin{equation}
 \mathbb P_{\rm cl}(E)\coloneqq
 \frac{\one_{\{\partial E=\varnothing\}}e^{-2\beta J|E|}}
      {\sum_{E':\,\partial E'=\varnothing}e^{-2\beta J|E'|}}.
 \label{eq:closed-defect-law}
\end{equation}
Here the subscript ``cl'' denotes closed sets. This auxiliary distribution keeps local closure and the same energy weights, but allows closed sets that do not satisfy the global boundary constraint. Equation~\eqref{eq:flux-law} gives exactly
\begin{equation}
 \mathbb P_{\rm phys}=\mathbb P_{\rm cl}
       (\,\cdot\mid E\text{ is a boundary}).
 \label{eq:physical-conditioning}
\end{equation}

Fix a closed connected set $\gamma$ of $j$ cells and restrict to configurations in which it is a connected component. Deleting this fixed $\gamma$ preserves closure and is injective: adding it back recovers the input. The input weight is $e^{-2\beta Jj}$ times the remaining weight. The sum of remaining weights is at most the full denominator in Eq.~\eqref{eq:closed-defect-law}, so
\begin{equation}
 \mathbb P_{\rm cl}(\gamma\text{ is a connected component})
       \le e^{-2\beta Jj}.
 \label{eq:component-deletion-bound}
\end{equation}
There are $n_{m+1}$ choices of a fixed dual cell. Summing the bound over these cells and all possible components, with harmless overcounting, gives for every integer $r\ge1$
\begin{equation}
 \begin{aligned}
 \mathbb P_{\rm cl}(\exists\gamma:|\gamma|\ge r)
 &\le n_{m+1}\sum_{j\ge r}D_{d,m}^{2(j-1)}e^{-2\beta Jj}\\
 &\le Q_L(r).
 \end{aligned}
 \label{eq:closed-peierls-tail}
\end{equation}
Here $\tau\coloneqq2\beta J-2\log D_{d,m}>0$, and the geometric-series bound is
\begin{equation}
 Q_L(r)\coloneqq n_{m+1}\frac{e^{-\tau r}}{1-e^{-\tau}}.
 \label{eq:peierls-Q}
\end{equation}
If every component has fewer than $c_{d,m}L$ cells, each has a local filling. Adding these fillings modulo two gives a filling of their union, so $E$ is a boundary. Consequently,
\begin{equation}
 \mathbb P_{\rm cl}(E\text{ is a boundary})
 \ge1-Q_L(\lceil c_{d,m}L\rceil).
 \label{eq:boundary-probability}
\end{equation}
At fixed $\tau>0$, this lower bound tends to one exponentially fast, up to the volume factor $n_{m+1}$. For sufficiently large $L$ it is at least $1/2$. Conditioning in Eq.~\eqref{eq:physical-conditioning} then increases any event probability by at most a factor of two, giving
\begin{equation}
 \mathbb P_{\rm phys}(\exists\gamma:|\gamma|\ge r)\le2Q_L(r).
 \label{eq:peierls-tail}
\end{equation}

\subsubsection{Local correction and its error}

Choose a constant $a_0>0$ and $r\coloneqq\lceil a_0\log L\rceil$, which is smaller than $c_{d,m}L$ for sufficiently large $L$. Starting from an occupied cell, inspect adjacent cells in the original defect set. Continue through occupied neighbors until either all neighbors of the discovered cells have been checked or $r$ occupied cells have been found. Correct only the first case with fewer than $r$ cells: it identifies an entire connected component. A fragment cut from a larger component need not be closed, so the local-filling argument does not apply to it.

All inspections use the original defect set, which is kept unchanged until the corrections are applied. Apply each component's correction once, using its first cell in a fixed ordering as its representative. Each output cell needs information only within distance $O(r)$; coloring overlapping output blocks gives $O(r^d)$ layers of gates with diameter $O(r)$.

When all components are smaller than $r$, the correction maps each defect fiber bijectively onto the full flat set. The unperturbed Gibbs law is uniform within each fiber, so its corrected output on this event is exactly $\rho_{\rm flat,L}$. On the complementary event, the trace-norm distance between any output state and $\rho_{\rm flat,L}$ is at most two. Multiplying this bound by the bad-event probability in Eq.~\eqref{eq:peierls-tail}, the decoder $\mathcal D_{r,L}^{\rm flux}$ satisfies
\begin{equation}
 \norm{\mathcal D_{r,L}^{\rm flux}(\rho_{\beta,0,L})
          -\rho_{\rm flat,L}}_1\le4Q_L(r).
 \label{eq:decoder-error}
\end{equation}
Taking $a_0>(d+s)/\tau$ gives error $O(L^{-s})$. This proves one conversion direction in an explicit sufficient domain $\beta J>\log D_{d,m}$. For $(d,m)=(3,1)$, $D_{3,1}=10$ and $\log D_{3,1}\simeq2.302585$. This decoder removes defects. To prove two-way equivalence, we must also sample and reinsert them. The next subsection constructs both directions in the presence of a weak classical perturbation.

\subsection{Two-way conversion with a weak classical perturbation}

\begin{figure}[!tbp]
\centering
\includegraphics[width=233.17651pt,height=80.98123pt]{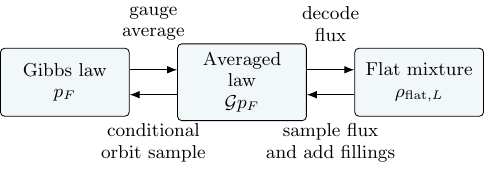}
\caption{The classical conversion has two distinct parts.
Uniform local gauge transformations map the physical Gibbs density
$p_F$ to $\mathcal Gp_F$. Sampling the conditional gauge law $Q_a$
of Eq.~\eqref{eq:orbit-conditional} gives the reverse conversion on
this state. Flux decoding then removes small components from the
averaged density; the reverse channel samples components and adds
their fillings to the flat input. Arrows denote state conversions
with polylogarithmic causal radius and vanishing whole-system
trace-norm error in the classical low-temperature domain.
The flat input supplies the winding information in the reverse direction.}
\label{fp:fig:route}
\end{figure}

Write the classical density as
\begin{equation}
 p_F(a)\coloneqq Z_F^{-1}
 e^{\beta J\sum_p B_p(a)+F(a)},\qquad F\coloneqq-\beta gV,
 \label{eq:classical-density}
\end{equation}
where $Z_F$ normalizes the sum over all spin configurations. Figure~\ref{fp:fig:route} separates the two operations needed for conversion: average and restore the gauge variables, and remove and restore the defects. We establish these operations in four steps. The probability law $p_F$ and its diagonal density matrix are identified throughout.

\subsubsection{Step 1: Gauge averaging and conditional recovery}

For an $(m-1)$-cell $u$, choosing $\eta_u=-1$ in Eq.~\eqref{eq:gauge-action} flips all physical $m$-cells containing $u$. To express the same transformation in bit notation, define $\widehat\eta_u\coloneqq(1-\eta_u)/2\in\{0,1\}$. The combined action is $a\oplus\delta\widehat\eta$, where $\oplus$ is addition modulo two and
\begin{equation}
 (\delta\widehat\eta)_\ell\coloneqq
          \sum_{u\in\partial\ell}\widehat\eta_u\pmod2.
 \label{eq:gauge-bits}
\end{equation}
Uniform independent gauge signs give a uniform average over the bits $\widehat\eta$. Different choices can have the same action, but their multiplicity is constant, and the gauge average is a bounded-depth local channel. Its exact use below relies on the local gauge invariance of $H_0$, with no corresponding requirement on $V$. The average acts on the perturbing Boltzmann weight $e^F$, not on $V$ itself.

Applying gauge averaging to the Boltzmann weight~\cite{FoersterNielsenNinomiya1980,Thorngren2026} gives the averaged perturbation weight and law
\begin{align}
 w_F(a)&\coloneqq2^{-n_{m-1}}\sum_{\widehat\eta}
                      e^{F(a\oplus\delta\widehat\eta)},\label{eq:gauge-weight}\\
 (\mathcal Gp_F)(a)&\propto e^{\beta J\sum_pB_p(a)}w_F(a).
 \label{eq:gauge-average}
\end{align}

The original law is recovered by sampling the gauge variables conditionally on the averaged configuration. Given $a$ drawn from $\mathcal Gp_F$, sample
\begin{equation}
 Q_a(\widehat\eta)\coloneqq
 \frac{e^{F(a\oplus\delta\widehat\eta)}}{2^{n_{m-1}}w_F(a)}
 \label{eq:orbit-conditional}
\end{equation}
and output $a\oplus\delta\widehat\eta$. In the joint weight $(\mathcal Gp_F)(a)Q_a(\widehat\eta)$, the factors $w_F(a)$ cancel. The remaining weight is $2^{-n_{m-1}}p_F(a\oplus\delta\widehat\eta)$, because the gauge transformation preserves every $B_p$. Summing over the $2^{n_{m-1}}$ possible gauge labels at fixed output therefore returns $p_F$ exactly.

This recovery can be implemented locally when $\beta|g|$ is small: the conditional distribution is then a weakly interacting model of the gauge variables, uniformly in the fixed input $a$. A heat-bath update resamples one gauge bit from its conditional law with the other bits fixed. A coloring groups updates with disjoint neighborhoods. Repeating these local updates samples $Q_a$ to inverse-power global accuracy in polylogarithmic depth. Thus conditional sampling reverses $\mathcal G$ on the specified Gibbs state, to the required accuracy.

\subsubsection{Step 2: Reduction to an interacting defect gas}

To decode defects, we rewrite the gauge-averaged weight $w_F$ as an interaction among defect cells. Expand each local function in products of cell signs. Gauge averaging retains products whose combined cell set is closed. A connected retained product with diameter much smaller than $L$ bounds an $(m+1)$-cell set, since $m\ge1$. Its sign product is therefore a product of the $B_p$ over a local filling. A convergent connected expansion gives
\begin{equation}
 \log w_F(a)=c_L-U(E(a))+r_L(a).\label{eq:effective-flux-expansion}
\end{equation}
Here $c_L$ is independent of $a$ and cancels upon normalization. The remainder $r_L(a)$ contains contributions with wrapping generating supports. Define the dimensionless potential
\begin{equation}
 U(E)\coloneqq\sum_X\Psi_X(E|_X).
\end{equation}
It adds to the defect cost $2\beta J|E|$. Here $X$ is a set of dual defect cells, $E|_X$ records their occupations, and each $\Psi_X$ is defined for arbitrary assignments. Although $U$ may be extensive, the expansion bounds its local interaction strength and exponential tail:
\begin{align}
 \sup_{p^\ast}\sum_{X\ni p^\ast}|X|e^{\mu\diam X}
               \norm{\Psi_X}_\infty&\le C\beta|g|,\\
 \norm{r_L}_\infty&\le Cn\beta|g|e^{-cL}.
 \label{eq:effective-flux}
\end{align}
Here $p^\ast$ ranges over dual defect cells, $\mu>0$ is fixed, and $C,c>0$ depend only on the fixed local data and $\mu$. SM Sec.~II proves a weighted version of Eqs.~\eqref{eq:effective-flux-expansion}--\eqref{eq:effective-flux}, also needed for quantum perturbations.

Dropping $r_L$ changes the full state by a small trace-norm error. The resulting defect law has weight $e^{-2\beta J|E|-U(E)}$ on sets $E$ that are boundaries. Deleting a closed component preserves closure but can violate this boundary constraint. As in Eqs.~\eqref{eq:closed-defect-law}--\eqref{eq:physical-conditioning}, we therefore use the auxiliary law
\begin{equation}
 \pi(E)\propto\one_{\{\partial E=\varnothing\}}
                         e^{-2\beta J|E|-U(E)}.
 \label{eq:interacting-flux}
\end{equation}
Conditioning $\pi$ on $E$ being a boundary recovers the physical defect law with potential $U$.

\subsubsection{Step 3: Local decoding to the flat ensemble}

Deleting a component changes $U$ by at most a small constant times its size. The energy cost $2J|\gamma|$ still suppresses large components, so the Peierls estimate and local fillings survive after reducing the allowed perturbation. Condition on the physical boundary constraint, discard the negligible probability of components larger than the logarithmic cutoff, and truncate the exponentially decaying interaction at logarithmic range. Write $U_r(E)\coloneqq\sum_{X:\diam X\le r}\Psi_X(E|_X)$ for the potential with logarithmic range cutoff $r$. These changes incur a small whole-state error; they are estimates of the target distribution, not postselection in the channel.

For the resulting law, the fixed filling map gives independent variables: a defect configuration drawn from the cutoff gas and a uniformly distributed flat configuration. The decoder measures and removes the former, leaving the latter. As in the unperturbed construction, it acts on every input and leaves large components untreated. Gauge averaging followed by this local decoder proves $p_F\to\rho_{\rm flat}$.

\subsubsection{Step 4: Defect sampling and restoration of the perturbation}

For the reverse direction, the input flat state supplies the winding information. Independent local random coins generate only the dilute defect gas. A component birth-and-death sampler, of the type used in loss-network constructions~\cite{FernandezFerrariGarcia2002}, deletes each component at rate one and inserts a compatible component $\gamma$ at rate
\begin{equation}
 b_\gamma(E)\coloneqq
 e^{-2\beta J|\gamma|-U_r(E\cup\gamma)+U_r(E)}.
 \label{eq:component-birth}
\end{equation}
Compatibility means that distinct components neither overlap nor touch along a face. Detailed balance gives the desired cutoff law. At sufficiently low temperature the total insertion influence is small, and a coupling contracts the total size of unmatched components at a size-independent rate. Running for $O(\log L)$ time therefore gives the desired whole-system accuracy.

The sampler is local after both component sizes and potential ranges are cut off at $O(\log L)$. Assign each candidate component to its first cell in a fixed ordering, called its root. Independent Poisson event records at these roots specify proposed insertions and deletions. The probability of a very long chronological chain of dependent events is bounded by an ordered-time factorial; truncating these chains at polylogarithmic length gives a finite local circuit. The circuit always produces an output, including on the rare bad records.

Adding the sampled fillings to the flat input restores the averaged density, and the conditional sampler of Eq.~\eqref{eq:orbit-conditional} restores the original perturbation. These steps prove the classical part of Theorem~\ref{thm:main} in both directions.

\section{Quantum perturbations and coherence recovery}
\label{sec:quantum}

For a quantum perturbation, we first reduce the diagonal probabilities to a weak classical interaction, then identify the conditional pure states whose mixture gives the full Gibbs state, and finally construct a local channel that prepares those states. This separates the defect conversion from the additional task of restoring coherences.

\subsection{Step 1: Diagonal reduction and classical conversion}

\begin{figure*}[t]
\centering
\includegraphics[width=444.30435pt,height=72.11092pt]{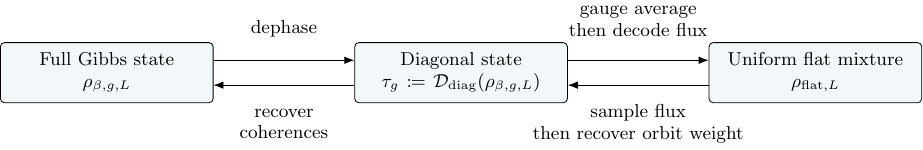}
\caption{Quantum low-temperature conversion through the exact diagonal
state $\tau_g$. Onsite $Z$-basis dephasing gives the forward arrow on
the left; conditional coherence recovery gives the reverse arrow.
The right-hand arrows contain both parts of the classical construction
in Fig.~\ref{fp:fig:route}, applied to the effective classical interaction
$F=\log\mathcal W_{\beta,g}$: gauge averaging followed by flux decoding,
or flux sampling and filling followed by conditional orbit sampling.
Every arrow denotes a conversion of the displayed state family up to
vanishing whole-system trace-norm error, with polylogarithmic causal
radius, in the quantum low-temperature domain. These channels need
not invert one another on arbitrary inputs. The flat input retains
the winding information needed by the reverse construction.}
\label{fp:fig:quantum-route}
\end{figure*}

Dephasing a noncommuting Gibbs state produces a classical probability law, but its effective interaction is not obtained by simply dropping the off-diagonal terms in $V$. Let $\mathcal D_{\rm diag}$ be onsite $Z$-basis dephasing and define
\begin{align}
 \tau_g&\coloneqq\mathcal D_{\rm diag}(\rho_{\beta,g,L}),\\
 \mathcal W_{\beta,g}(a)&\coloneqq e^{\beta H_0(a)}\bra a e^{-\beta H_g}\ket a.
\label{eq:quantum-diagonal}
\end{align}
Figure~\ref{fp:fig:quantum-route} shows the resulting conversion route. The state $\tau_g$ is an intermediate diagonal ensemble, not the Gibbs state obtained by keeping only the diagonal part of $V$.

The diagonal density is proportional to
$e^{-\beta H_0(a)+\log \mathcal W_{\beta,g}(a)}$. The exact interaction-picture expansion involves
\begin{equation}
 v_X(u)\coloneqq e^{u\beta H_0}v_Xe^{-u\beta H_0},
 \qquad 0\le u\le1.
 \label{eq:dressed-term}
\end{equation}
Here $u$ is dimensionless imaginary time. Only interaction cells meeting $X$ contribute to this conjugation. Since one physical cell belongs to $2(d-m)$ interaction cells,
\begin{equation}
 \norm{v_X(u)}\le e^{4b(d-m)\beta J},\qquad
 \lambda\coloneqq\beta|g|e^{4b(d-m)\beta J}.
 \label{eq:lambda}
\end{equation}
Each dressed term still has bounded support, contained in the union of $X$ and the supports $\partial p$ of all $B_p$ with $\partial p\cap X\ne\varnothing$. We call this union its enlarged support.

The effective classical interaction is obtained by grouping the imaginary-time terms by overlapping enlarged supports. Disjoint groups factor in the product basis, and the logarithm retains connected groups. Their total local strength is $O(\lambda)$ even after weighting each contribution exponentially in its number of generating interaction labels. This bound controls the resulting long-range tails, so the weighted classical construction applies; SM Sec.~III gives the estimates. The logarithm is real because $\mathcal W_{\beta,g}(a)>0$. It follows that $\tau_g\sim_{\LC}\rho_{\rm flat}$ when $\lambda$ is small and the temperature is sufficiently low.

\subsection{Step 2: Conditional pure-state representation}

Restoring the off-diagonal matrix elements is a distinct problem. Turn on the perturbation with a dimensionless path parameter $0\le t\le1$, keeping $\beta$ fixed, and define
\begin{align}
 H(t)&\coloneqq H_0+tgV,\label{eq:conditional-path}\\
 M_t&\coloneqq e^{-\beta H(t)/2},\\
 d_t(a)&\coloneqq\bra a M_t^2\ket a,\\
 \ket{\psi_{t,a}}&\coloneqq\frac{M_t\ket a}{\sqrt{d_t(a)}},\\
 p_t(a)&\coloneqq\frac{d_t(a)}{\Tr M_t^2}.
 \label{eq:conditional-vectors}
\end{align}
Completeness of the basis gives the exact product-basis thermal-ensemble identity~\cite{White2009}
\begin{equation}
 \frac{M_t^2}{\Tr M_t^2}
 =\sum_a p_t(a)\proj{\psi_{t,a}}.
 \label{eq:conditional-ensemble}
\end{equation}
At $t=0$, $\ket{\psi_{0,a}}=\ket a$ because $H_0$ is diagonal. At $t=1$, $\tau_g=\sum_a p_1(a)\proj a$ already supplies the required probabilities. We need a deterministic channel that maps each label $a$ to its conditional vector, while preserving that probability distribution.

\Needspace{12\baselineskip}
\subsection{Step 3: Local coherence recovery}

We construct a positive auxiliary Hamiltonian whose unique zero-energy state is the desired conditional vector, adapting the thermal-parent construction~\cite{FeiguinKlich2013}:
\begin{align}
 P_{\ell}(a)&\coloneqq I-\proj{a_{\ell}},\\
 Q_{\ell}(t,a)&\coloneqq M_tP_{\ell}(a)M_t^{-1},\\
 \mathcal K_a(t)&\coloneqq\sum_{\ell}Q_{\ell}(t,a)^\dagger Q_{\ell}(t,a).
 \label{eq:auxiliary-parent}
\end{align}
Here $P_\ell(a)$ rejects the specified onsite bit. The common zero space of all its conjugates $Q_\ell$ is the span of $M_t\ket a$. Thus $\mathcal K_a(t)$ has this one-dimensional kernel, while
$\mathcal K_a(0)=\sum_{\ell}P_{\ell}(a)$ has a unique product ground state and gap one. Connected commutator estimates show that $\mathcal K_a(t)$ differs from this product parent by terms that decay exponentially in their range and are small in a local interaction norm, uniformly in $a,t,L$. The product-state case of gap stability~\cite{BravyiHastings2011} then gives a uniform auxiliary gap. This is not a gap assumption on $H_g$.

Attach an onsite control register carrying a copy of $a$ and combine the parents as a controlled, local Hamiltonian. Spectral flow is a unitary evolution transporting the zero-energy space along a uniformly gapped Hamiltonian path~\cite{BachmannEtAl2012}. Here it maps
\begin{equation}
 \ket a_{\rm C}\ket a_{\rm P}
 \longmapsto e^{i\theta_a}\ket a_{\rm C}\ket{\psi_{1,a}}_{\rm P}.
 \label{eq:controlled-flow}
\end{equation}
Here C labels the control registers, P the physical spins, and $\theta_a$ is an irrelevant real phase. The controls keep different labels orthogonal even though the conditional physical vectors need not be orthogonal. The flow preserves every control label.

Almost-exponential decay of the flow generator and a local block decomposition~\cite{HaahEtAl2021} give a polylogarithmic-radius circuit with any prescribed inverse-power operator-norm error. Copying the diagonal input, applying that circuit, and discarding the controls therefore restores $\rho_{\beta,g,L}$ with the required trace error. Imaginary time defines the target vectors; the physical preparation is the deterministic local circuit, with no filtering or postselection.

Composing dephasing with the classical decoder gives the forward channel. Composing the classical encoder with coherence recovery gives the reverse channel. Their errors and causal radii add. Tightening the constants to cover the weighted classical expansion and auxiliary gap gives the quantum window in Eq.~\eqref{eq:low-domain}.

\section{High temperature and physical ancillary systems}
\label{sec:high-ancilla}
The high-temperature preparation starts from local product inputs and does not use flux geometry. We then include independent onsite degrees of freedom as part of the physical target state, distinguishing them from disposable work registers.

\subsection{Preparation from a product state}

Local-channel preparation under uniform clustering and approximate Markov assumptions on all induced subsystems was established in Ref.~\cite{BrandaoKastoryano2019}. General local quantum Hamiltonians also admit high-temperature Gibbs samplers~\cite{RouzeFrancaAlhambra2026Thermalization,RouzeFrancaAlhambra2026Optimal}, including implementations by local circuits~\cite{HahnEtAl2026}. Here an auxiliary-parent construction gives the geometric locality and whole-system accuracy required by our channel convention.

Write $H_g=\sum_Yh_Y$ with support size at most
$b'\coloneqq\max\{2(m+1),b\}$ and fixed diameter. The local energy scale is
\begin{equation}
 j_g\coloneqq2(d-m)J+n_{\mathrm{inc}}|g|,\qquad
 \sup_{\ell}\sum_{Y\ni\ell}\norm{h_Y}\le j_g.
 \label{eq:local-energy}
\end{equation}

To prepare a purification of the Gibbs state, attach one work qubit to each physical qubit. The normalized onsite Bell pair is $\ket{\Omega_\ell}\coloneqq(\ket{00}_\ell+\ket{11}_\ell)/\sqrt2$, with the first entry physical and the second its work-register partner. Prepare their product,
\begin{equation}
 \ket\Omega\coloneqq\bigotimes_{\ell}\ket{\Omega_\ell}.
 \label{eq:bell-reference}
\end{equation}
The normalized vector proportional to
$(e^{-t\beta H_g/2}\otimes I)\ket\Omega$ purifies the Gibbs state at inverse temperature $t\beta$. Its auxiliary parent starts from the sum $\sum_\ell(I-\proj{\Omega_\ell})$, which has the product of Bell pairs as its unique zero-energy state and gap one. A nested-commutator expansion is small when $\beta b'j_g$ is small. The same product-gap and spectral-flow construction prepares the endpoint purification locally; discarding the work registers prepares the Gibbs state. Exact onsite depolarization is the reverse channel. This proves Eq.~\eqref{eq:high-equivalence}, as detailed in SM Sec.~IV.

For the unperturbed classical model, a more elementary explicit bound is available. One cell participates in $2(d-m)$ interactions, each containing $2m+1$ other spins. The influence sum $\alpha$ adds, over neighboring spins, the largest total-variation change of the single-spin conditional law when that neighbor is flipped. It obeys
\begin{equation}
 \alpha\le2(d-m)(2m+1)\tanh(\beta J).
 \label{eq:dobrushin-unperturbed}
\end{equation}
When the right side is below one, a complete colored sweep contracts the maximum disagreement probability by at least this factor. After $t$ sweeps, the whole-system trace error is at most $2n\alpha^t$. Thus $t=O(\log L)$ suffices. In particular,
\begin{equation}
 \beta J<\operatorname{arctanh}
       \frac1{2(d-m)(2m+1)}
 \label{eq:explicit-high}
\end{equation}
is a concrete product-equivalence domain. For $(d,m)=(3,1)$, its upper threshold is $\operatorname{arctanh}(1/12)\simeq0.083527$. This boundary, like the low-temperature Peierls bound, is not a prediction of the transition temperature.

\subsection{Ancillas included in the Hamiltonian}

Disposable work registers should be distinguished from extra physical degrees of freedom whose joint Gibbs state is part of the target. Suppose a fixed finite number of bounded-dimensional ancillary sites per lattice cell have independent onsite Hamiltonians,
\begin{equation}
 H_A\coloneqq\sum_jh_j,\qquad
 \tau_A\coloneqq\bigotimes_j\frac{e^{-\beta h_j}}{\Tr e^{-\beta h_j}}.
 \label{eq:physical-ancillas}
\end{equation}
Let $\Delta_A$ bound their spectral widths. The joint Hamiltonian is $H_0+H_A+gV$, with identity factors understood. The terms in $gV$ may include system--ancilla and ancilla--ancilla couplings, with the same fixed locality conventions on the joint sites. The low-temperature reference becomes $\rho_{\rm flat,L}\otimes\tau_A$.

For the classical branch, fix an onsite eigenbasis of each $h_j$ and require $V$ to be diagonal in its tensor product with the system $Z$ basis. Write $F(a,z)\coloneqq-\beta gV(a,z)$, where $z$ labels these ancillary eigenstates. Summing over $z$ with the product weights $\tau_A(z)$ gives the system exponent $\log\mathbb E_{z\sim\tau_A}e^{F(a,z)}$. Product expectation preserves the weighted expansion bounds. The conditional ancillary distribution is then restored by a weak-interaction sampler, uniformly even when some onsite probabilities are small. For quantum couplings, conjugation by $H_A$ contributes a factor exponential in $\beta\Delta_A$ to the imaginary-time estimate.

\begin{theorem}[Stability with physical ancillary systems]
\label{thm:ancilla}
For $1\le m\le d-2$ and the joint model just specified, there are positive class-dependent constants $\tilde B_A,\tilde a_A,\tilde C_A,\tilde a'_A$ such that the sufficient low-temperature windows
\begin{equation}
 \beta J>\tilde B_A,
 \label{eq:ancilla-temperature}
\end{equation}
\begin{equation}
 \begin{cases}
 \beta|g|<\tilde a_A,&\text{classical},\\
 \beta|g|e^{\tilde C_A\beta(J+\Delta_A)}<\tilde a_A,
                  &\text{Hermitian}.
 \end{cases}
 \label{eq:ancilla-window}
\end{equation}
imply two-way conversion of the full joint state to $\boldsymbol\rho_{\rm flat}\otimes\boldsymbol\tau_A$. At high temperature,
\begin{equation}
 \beta[2(d-m)J+\Delta_A/2+n_{\mathrm{inc}}|g|]<\tilde a'_A
 \label{eq:ancilla-high-window}
\end{equation}
implies two-way equivalence to an onsite product state, with fixed site multiplicities absorbed into the constants. All conversions have product work-register inputs, polylogarithmic causal radius, and whole-system trace-norm error $O(L^{-s})$ for each fixed $s>0$.
\end{theorem}

No gap or nondegeneracy of $h_j$ is required. Interacting ancillary reference states are outside this extension. SM Sec.~IV proves the claim for the full joint density matrix.

\section{A winding-correlation invariant}
\label{sec:invariant}
The winding invariant measures information that two distant extended regions can read in common. Its monotonicity under local conversion turns the reference values into a phase distinction.

\subsection{Reader geometry and ordered limits}

\begin{figure}[t]
\centering
\includegraphics[width=202.66776pt,height=84.40530pt]{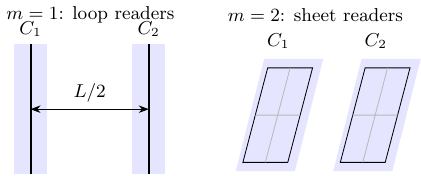}
\caption{Reader geometry for the invariant $\mathcal F_m$.
Each $C_i$ winds in $m$ fixed coordinate directions; its surrounding
region includes all physical cells within transverse distance $\zeta L$.
Opposite ends of each line, and opposite edges of each sheet, are identified.
The illustrated tori $C_i$ are separated transversely by $L/2$, corresponding to $\kappa=1/2$.
The stable $m=2$ models have $d\ge4$; the shaded sheets schematically represent thick neighborhoods
in the additional transverse directions. In the flat ensemble, products
of $Z_\ell$ over $C_1$ and $C_2$ read the same unbiased binary label.
The invariant optimizes over all binary observables in these regions.}
\label{fig:readers}
\end{figure}

Fix a coordinate orientation, say the first $m$ axes. Let $C_{1,L}$ and $C_{2,L}$ be two parallel primitive coordinate $m$-tori, each wrapping once around those axes, and separated transversely by at least $\kappa L$ for a fixed $\kappa>0$. For a dimensionless relative width $\zeta>0$, the reader region $R_{i,L}(\zeta)$ contains the physical $m$-cells within transverse distance $\zeta L$ of $C_{i,L}$, including cells intersecting the edge of this neighborhood. This last convention adds only a bounded number of lattice spacings. For sufficiently small fixed $\zeta$, the regions are disjoint, periodic in the first $m$ directions, and bounded in the transverse directions. The orientation, positions, and separation convention are fixed independently of the state (Fig.~\ref{fig:readers}).

\Needspace{9\baselineskip}
For binary Hermitian readers $A_i$, define
\begin{align}
 F_{\zeta,L}^{(m)}(\rho)&\coloneqq
 \sup_{\substack{A_i=A_i^\dagger,\ A_i^2=I\\
             \supp A_i\subset R_{i,L}(\zeta)}}
 \left|\Tr\rho A_1A_2-\Tr\rho A_1\Tr\rho A_2\right|,\label{eq:finite-reader}\\
 \mathcal F_m(\boldsymbol\rho)&\coloneqq
 \lim_{\zeta\downarrow0}\limsup_{L\to\infty}
                  F_{\zeta,L}^{(m)}(\rho_L).
 \label{eq:F-definition}
\end{align}
The thermodynamic limit is taken before the relative width tends to zero. The outer limit exists by monotonicity in the width, and Cauchy--Schwarz gives $0\le\mathcal F_m\le1$. The reader optimization allows arbitrary observables in each region. Equivalently, the readers can first perform local recovery, provided its causal radius is included in their total transverse-width budget. This is the higher-cell version of the winding-correlation construction of Ref.~\cite{Watanabe2026}. An optimized regional covariance also appears in Ref.~\cite{MaKhemaniSang2026}, where observables in the interaction algebra test clustering for an annulus partition. Here arbitrary binary readers on separated winding neighborhoods, with the ordered limits above, define a channel invariant.

\begin{proposition}[Monotonicity and two-way invariance]
\label{prop:F-monotone}
If $\boldsymbol\rho\to_{\LC}\boldsymbol\sigma$, then
$\mathcal F_m(\boldsymbol\sigma)\le\mathcal F_m(\boldsymbol\rho)$.
Consequently $\mathcal F_m$ is constant on every two-way local-channel equivalence class.
\end{proposition}

For fixed readers, changing the state by trace norm $\varepsilon$ changes the covariance by at most $3\varepsilon$. Pulling two readers through a conversion channel enlarges each region by $o(L)$. For fixed $\zeta$, their full backward cones remain disjoint and fit within $R_{i,L}(2\zeta)$ for large $L$. Product work-register inputs then give a product of two pulled-back Hermitian contractions. A Hermitian contraction is an observable with eigenvalues in $[-1,1]$. It is a convex combination of binary readers, so the optimized covariance cannot increase. Taking the ordered limits proves the proposition. This uses the disjoint local dilation, not multiplicativity of a general channel adjoint; SM Sec.~V supplies the details.

\subsection{Exact reference values and phase separation}

The Wilson products in Eq.~\eqref{eq:Wilson} on the two reader tori read the same unbiased sign in the flat ensemble; we suppress the size label $L$. At $m=1$ these are Wilson loops. The two parallel tori bound a connecting $(m+1)$-dimensional slab. Multiplying $B_p$ over that slab cancels every interior cell twice and leaves $W_{C_1}W_{C_2}$. In a flat configuration this product is one. Each individual $W_{C_i}$ has mean zero in the canonical flat ensemble: a wrapping $U_\Sigma$ on a dual complementary sheet intersects each torus once and reverses both signs, pairing configurations with equal weights. Therefore
\begin{equation}
 \langle W_{C_1}W_{C_2}\rangle_{\rm flat}=1,
 \qquad \langle W_{C_i}\rangle_{\rm flat}=0,
 \qquad \mathcal F_m(\boldsymbol\rho_{\rm flat})=1.
 \label{eq:F-flat}
\end{equation}
The universal upper bound one makes this an exact optimized value. Related obstructions based on separated logical readers occur in bounds on dissipative preparation of topological order~\cite{KonigPastawski2014}.

For a product input, the two regional reports are independent; disjoint backward cones preserve their independence under allowed local circuits. Hence
\begin{equation}
 \mathcal F_m(\boldsymbol\rho_\infty)=0.
 \label{eq:F-product}
\end{equation}
Theorems~\ref{thm:main} and~\ref{thm:high}, together with Proposition~\ref{prop:F-monotone}, give
\begin{equation}
 \boxed{\quad
 \mathcal F_m(\boldsymbol\rho_{\beta,g})=
 \begin{cases}
 1,&\text{low-temperature domains},\\
 0,&\text{high-temperature domain}.
 \end{cases}\quad}
 \label{eq:F-domains}
\end{equation}
Adding independent physical ancillary states does not change these values: append and discard them by onsite channels.

Bare Wilson surfaces can be sensitive to the many small defects met by an extended support. The optimized quantity retains the local correction inside the reader region, which is why Eq.~\eqref{eq:F-domains} follows from whole-state conversion. The unperturbed decoder alone also gives $\mathcal F_m=1$ in its explicit domain: monotonicity supplies a lower bound of one, already the maximum. That argument does not by itself supply the reverse channel or extend the result to an entire unknown low-temperature phase.

\section{Scope and outlook}
\label{sec:discussion}

The general cell model separates the two ingredients of symmetry-independent thermal information: a sector label that a bounded region cannot read and defects that can be filled locally when their components are small. Their intersection gives $1\le m\le d-2$. The canonical flat mixture supplies the nonlocal random information; the construction transports that information while locally removing and restoring fluctuations. It never generates a winding sign from independent work-register inputs.

The exact defect constraint also distinguishes this mechanism from classical pyrochlore spin ice. There, the ice rule is energetic: thermal monopoles permit string endpoints and act as dual ordering fields, rounding continuous transitions of the monopole-free model~\cite{WatanabeMotomeWatanabeSpinOne2026,WatanabeMotomeWatanabeSpinS2026}. Here Eq.~\eqref{eq:bianchi} is an operator identity for any $V$: simultaneous $B_p$ measurements still give closed defect sets, even for noncommuting perturbations. Closure excludes this endpoint mechanism, but local-channel stability additionally requires locally indistinguishable sectors and controlled defect statistics; closure alone does not prove a thermodynamic singularity.

The conclusions concern full density matrices, with errors that vanish globally, and they survive weak noncommuting perturbations through an explicit coherence-recovery channel. The auxiliary gap used for that channel is a preparation tool, not a statement about the spectrum or relaxation time of the physical model. The reference states are classical, and the invariant reads a commuting bit. Neither a quantum-memory lifetime nor recovery of a noncommuting logical algebra is established.

The companion Letter~\cite{WatanabeThermodynamics2026} establishes free-energy analyticity separately, assuming a fixed finite spatial period. The channel results here require no such periodicity. Tildes distinguish the constants in the present state-level estimates from the Letter's thermodynamic estimates. A common sufficient low-temperature domain is obtained by taking the larger of the two temperature thresholds, the smaller coupling threshold, and, for quantum perturbations, the larger exponential constant. At high temperature, take the smaller of the two sufficient thresholds. Thus analyticity and local-channel equivalence hold simultaneously on common domains without identifying their boundaries with a critical line.

The sufficient temperature windows are deliberately conservative. We do not determine their sharp boundaries, the number or order of intervening thermodynamic transitions, or two-way equivalence at every point where the unperturbed model has low-temperature order. In particular, a special 3D duality or field argument cannot settle these questions for all cell degrees. Extending the local transport construction beyond the dilute-defect regime, while retaining global accuracy and control of winding information, remains a separate problem.
\label{last-main-page}
\begin{acknowledgments}
We thank Hal Tasaki for helpful discussions.
OpenAI Codex was used under the authors' direction to assist with drafting, organizing and checking derivations, and preparing schematic figures.
\end{acknowledgments}
\section*{Data availability}
No data were generated in this study.
% All works cited only in the SM are also listed here, as requested by APS.

% ===== Main bibliography =====

% ===== Supplemental Material: independent S numbering and citation namespace =====
\onecolumngrid
\clearpage
\setcounter{section}{0}
\setcounter{subsection}{0}
\setcounter{subsubsection}{0}
\setcounter{equation}{0}
\setcounter{figure}{0}
\setcounter{table}{0}
\setcounter{theorem}{0}
\renewcommand{\theequation}{S\arabic{equation}}
\renewcommand{\thefigure}{S\arabic{figure}}
\renewcommand{\thetable}{S\arabic{table}}
\renewcommand{\thetheorem}{S\arabic{theorem}}
\providecommand{\theHsection}{}
\providecommand{\theHsubsection}{}
\providecommand{\theHsubsubsection}{}
\providecommand{\theHequation}{}
\providecommand{\theHfigure}{}
\providecommand{\theHtable}{}
\providecommand{\theHtheorem}{}
\renewcommand{\theHsection}{SM.\arabic{section}}
\renewcommand{\theHsubsection}{SM.\arabic{section}.\arabic{subsection}}
\renewcommand{\theHsubsubsection}{SM.\arabic{section}.\arabic{subsection}.\arabic{subsubsection}}
\renewcommand{\theHequation}{SM.\arabic{equation}}
\renewcommand{\theHfigure}{SM.\arabic{figure}}
\renewcommand{\theHtable}{SM.\arabic{table}}
\renewcommand{\theHtheorem}{SM.\arabic{theorem}}
\phantomsection
\pdfbookmark[0]{Supplemental Material}{supplement-start}
\begin{center}
{\large\bfseries Supplemental Material:\\[3pt]
Local-Channel Stability of Classical $\mathbb Z_2$ Lattice Gauge Models\par}
\medskip
Guanxi Xun,$^{1}$ Weiguang Cao,$^{1,2}$ and Haruki Watanabe$^{1,2,3}$\par
\smallskip
{\small\itshape
$^1$Department of Physics, Hong Kong University of Science and Technology, Clear Water Bay, Hong Kong, China\par
$^2$Center for Theoretical Condensed Matter Physics, Hong Kong University of Science and Technology, Clear Water Bay, Hong Kong, China\par
$^3$Institute for Advanced Study, Hong Kong University of Science and Technology, Clear Water Bay, Hong Kong, China\par}
\smallskip
\today
\end{center}
\medskip

\noindent This Supplemental Material gives the complete proofs of local-channel stability and the winding-correlation invariant. It includes the definitions and statements used in the proofs, so that it can be read independently of the main text. Section~\ref{sm:sec:model} states the common setup and low-temperature theorem. Section~\ref{sm:app:classical} proves its classical part, and Sec.~\ref{sm:app:quantum} proves its quantum part by diagonal reduction and coherence recovery. Section~\ref{sm:app:high-ancilla} treats high temperature and independent physical ancillary systems, and Sec.~\ref{sm:app:invariant} proves the reader invariant.

\section{Model, locality, and low-temperature statement}
\label{sm:sec:model}

We fix the model, geometric conventions, and allowed channels before stating the low-temperature theorem for both classical and quantum perturbations.

Cell-set addition and incidence counts use binary parity, with the spin signs given by $\sigma_\ell=(-1)^{a_\ell}$. A configuration can equivalently be identified with the set of cells carrying bit one, so bitwise addition is symmetric difference. Sums of real coefficients and probabilities retain their usual meaning.

Fix a periodic hypercubic lattice of side $L$, spatial dimension $d\ge1$, and cell dimension $0\le m\le d-1$. The low-temperature arguments use $1\le m\le d-2$. Let $\Lambda_L^{(j)}$ be its elementary $j$-cells, and set $N\coloneqq L^d$, $n_j\coloneqq\binom djN$, and $n\coloneqq n_m$. A physical qubit occupies each $m$-cell $\ell$, with Pauli operators $Z_\ell,X_\ell$. For an $(m+1)$-cell $p$, let $\partial p$ be its $2(m+1)$ boundary $m$-cells and define
\begin{align}
 B_p&\coloneqq\prod_{\ell\in\partial p}Z_\ell,\qquad
 H_0\coloneqq H_{d,m}\coloneqq-J\sum_pB_p,\qquad J>0,\label{sm:eq:model}\\
 H_g&\coloneqq H_0+gV,\qquad V\coloneqq\sum_Xv_X,\\
 \rho_{\beta,g,L}&\coloneqq Z_L(\beta,g)^{-1}e^{-\beta H_g},\qquad
 Z_L(\beta,g)\coloneqq\Tr e^{-\beta H_g}.\label{sm:eq:perturbed}
\end{align}
Here $\beta\coloneqq1/T$, and the lattice spacing and Boltzmann constant are one.

Each dimensionless Hermitian $v_X$ has norm at most one, assigned nonempty support $X$ of at most $b$ physical cells, and diameter at most $R$. Each physical cell belongs to at most $n_{\mathrm{inc}}$ assigned supports. These local data are fixed independently of $L$, and the local operators are independent of $\beta,g$. No spatial periodicity or internal symmetry of $V$ is required. A perturbation is classical when the total $V$ is diagonal in the product $Z$ basis. In that case, replacing each $v_X$ by its diagonal part leaves their sum $V$ unchanged and does not increase norms, supports, or incidence bounds. Every trace uses the full spin Hilbert space, without a Gauss-law constraint or a fixed winding sector. We write $\norm{O}$ for operator norm, $\norm{\rho-\sigma}_1$ for trace norm, and $\norm{f}_\infty$ for supremum norm; estimate constants depend only on fixed local data.

Write the spin signs as $\sigma_\ell\coloneqq(-1)^{a_\ell}$, where $a_\ell\in\{0,1\}$, and identify $\ket a$ with $\ket\sigma$. A configuration is flat if $B_p(a)=1$ for all $p$. With $\mathcal A_{\rm flat,L}$ the full set of flat configurations, the references are
\begin{equation}
 \rho_{\rm flat,L}\coloneqq\frac1{|\mathcal A_{\rm flat,L}|}
       \sum_{a\in\mathcal A_{\rm flat,L}}\proj a,
 \qquad \rho_{\infty,L}\coloneqq2^{-n}I.
 \label{sm:eq:references}
\end{equation}
The first mixes all winding sectors; the second is an onsite product state.

For $m\ge1$, independent signs $\eta_u$ on $(m-1)$-cells act by
\begin{equation}
 (\eta\sigma)_\ell\coloneqq\sigma_\ell\prod_{u\in\partial\ell}\eta_u.
 \label{sm:eq:gauge-action}
\end{equation}
With $\widehat\eta_u\coloneqq(1-\eta_u)/2$, the same action is $a\oplus\delta\widehat\eta$, where
\begin{equation}
 (\delta\widehat\eta)_\ell\coloneqq
   \sum_{u\in\partial\ell}\widehat\eta_u\pmod2.
 \label{sm:eq:gauge-bits}
\end{equation}
It preserves every $B_p$; gauge-related configurations remain distinct physical states.

In the low-temperature range, let $k\coloneqq d-m-1$ and $E(a)\coloneqq\{p^\ast:B_p(a)=-1\}$ be the occupied dual $k$-cells. Here $p^\ast$ is the dual cell crossing the physical interaction cell $p$. A cell-set boundary consists of faces of odd incidence, and cell-set addition is symmetric difference. The set $E(a)$ is a boundary, hence closed. Two dual $k$-cells are adjacent when they share a $(k-1)$-face. A connected component is a maximal set joined by these adjacencies, including every occupied cell reachable from any of its cells. The adjacency degree is at most $D_{d,m}\coloneqq2(d-m-1)(2m+3)$: a cell has $2k$ faces, each with $2(d-k+1)-1$ other incident cells.

For each connected set of $j$ cells containing a fixed cell, choose a spanning tree, namely $j-1$ adjacency links connecting the cells without a cycle. A fixed ordering makes the choice and traversal deterministic. Traversing each tree link twice from the fixed cell gives a walk of $2(j-1)$ steps with at most $D_{d,m}$ choices per step. Its visited cells recover the set. Hence at most $D_{d,m}^{2(j-1)}$ connected sets of size $j$ contain the fixed cell.

A defect fiber is the set of spin configurations giving the same $E$. Since $H_0(a)=-Jn_{m+1}+2J|E(a)|$ and all nonempty defect fibers have equal size, the unperturbed defect law is
\begin{equation}
 \mathbb P_{\rm phys}(E)\propto
 \one_{\{E\text{ is a boundary}\}}e^{-2\beta J|E|}.
 \label{sm:eq:flux-law}
\end{equation}
The indicator $\one_{\{\cdots\}}$ is one when the stated condition holds and zero otherwise. The law is normalized over all dual $k$-cell sets, with nonzero weight only for boundaries.

For a real perturbing exponent $F$, define the normalized classical density
\begin{equation}
 p_F(a)\coloneqq Z_F^{-1}e^{\beta J\sum_pB_p(a)+F(a)},
 \qquad F\coloneqq-\beta gV\ \text{for a classical perturbation}.
 \label{sm:eq:classical-density}
\end{equation}
Here $Z_F\coloneqq\sum_a e^{\beta J\sum_pB_p(a)+F(a)}$ normalizes the sum over all spin configurations. Probability laws and their diagonal density matrices are identified.

A physical local channel is a circuit of geometrically supported completely positive trace-preserving maps. Its work registers start in an input-independent product state over local sites. All classical messages and work registers count toward the causal radius, which bounds support expansion in a local dilation of the full circuit. No postselection or external shared randomness is allowed. Block gates may perform arbitrary operations inside their support, with no claim of efficient elementary-gate decomposition. For state families $\boldsymbol\rho\coloneqq\{\rho_L\}_L$ and $\boldsymbol\sigma\coloneqq\{\sigma_L\}_L$, write $\boldsymbol\rho\longrightarrow_{\LC}\boldsymbol\sigma$ when channels satisfy
\begin{equation}
 R_{\Phi_L}=o(L),\qquad
 \norm{\Phi_L(\rho_L)-\sigma_L}_1\longrightarrow0,
 \label{sm:eq:lc}
\end{equation}
and $\boldsymbol\rho\sim_{\LC}\boldsymbol\sigma$ when both directions exist. The channels need only convert the specified state families; their composition need not act as the identity on arbitrary inputs.
\begin{theorem}[Low-temperature local-channel stability]
\label{sm:thm:main}
Fix $d,m$ with $1\le m\le d-2$ and a local interaction class $(b,R,n_{\mathrm{inc}})$. There are positive constants $\tilde B,\tilde a,\tilde C$ such that either of the conditions
\begin{equation}
 \begin{cases}
 \beta J>\tilde B,\quad \beta|g|<\tilde a,
       &V\text{ classical},\\
 \beta J>\tilde B,\quad \beta|g|e^{\tilde C\beta J}<\tilde a,
       &V\text{ Hermitian}
 \end{cases}
 \label{sm:eq:low-domain}
\end{equation}
implies
\begin{equation}
 \boldsymbol\rho_{\beta,g}\sim_{\LC}\boldsymbol\rho_{\rm flat}.
 \label{sm:eq:low-equivalence}
\end{equation}
For every fixed $s>0$, each conversion has whole-system trace-norm error $O(L^{-s})$, polylogarithmic gate diameter, depth, and causal radius, and product work-register inputs.
\end{theorem}

Section~\ref{sm:app:classical} proves the weighted classical equivalence of Proposition~\ref{sm:prop:weighted}; its finite-range specialization gives the classical part of Theorem~\ref{sm:thm:main}. That proof is completed in Sec.~\ref{sm:app:reverse-sampling}. Section~\ref{sm:app:quantum} first reduces the diagonal quantum Gibbs state to the weighted classical problem, then constructs a local channel restoring its coherences. The quantum proof is completed in Sec.~\ref{sm:app:controlled-flow}.

Setting $g=0$ in Theorem~\ref{sm:thm:main} gives $\boldsymbol\rho_{\beta,0}\sim_{\LC}\boldsymbol\rho_{\rm flat}$. Composition then yields $\boldsymbol\rho_{\beta,g}\sim_{\LC}\boldsymbol\rho_{\beta,0}$ at the same temperature. Trace-norm contractivity bounds the composed error by the sum of the two errors; depths and causal radii also add. Using independent work registers for the two channels preserves product initialization, so all stated accuracy and locality bounds remain valid.

\section{Weighted classical transport and its local implementation}
\label{sm:app:classical}

This section proves the classical conversion used in the main text, including the exponentially decaying interactions produced by the quantum reduction. Throughout the low-temperature construction, $1\le m\le d-2$. In the conversion estimates, $0<\varepsilon<1$ denotes the target accuracy. For the classical construction we take $\varepsilon=L^{-s}$ with fixed $s>0$; logarithmic cutoffs are then smaller than the system size for sufficiently large $L$.

\subsection{Local fillings and physical defect fibers}
\label{sm:app:fillings}

Local fillings define the decoder and its reverse; we construct them explicitly for nonwrapping components of the torus. The elementary cubical sweep, local gauge exactness, and equal-fiber argument below are also used in the companion Letter's defect representation~\cite{SM-WatanabeThermodynamics2026}.

The boundary of a positive-dimensional cell set consists of its faces with odd incidence, using the primal or dual lattice as appropriate. For a primal $j$-cell set $S$, define its coboundary $\delta S$ as the $(j+1)$-cells incident on an odd number of cells in $S$. Equivalently, regard $S$ as a binary assignment: $\delta$ takes its sum around each $(j+1)$-cell. The assignment is flat when all these sums vanish modulo two. For $j\ge1$, the gauge map takes an assignment on $(j-1)$-cells to its boundary sums on $j$-cells; at $j=m$ this is Eq.~\eqref{sm:eq:gauge-bits}. The identities $\partial^2=0$ and $\delta^2=0$ mean empty boundary and empty coboundary, respectively.

Cubical duality reverses incidence: $\ell\in\partial p$ if and only if $p^\ast\in\partial\ell^\ast$. Flipping $\ell$ therefore toggles precisely the dual defect cells in $\partial\ell^\ast$. Thus the defect set of $a$ is the boundary of the dual of its occupied physical cells, proving the global constraint in Eq.~\eqref{sm:eq:flux-law}.

The physical filling rewrites a closed spin product as a product of $B_p$'s;
the dual filling gives the spin flips that remove or insert a defect
component. In three dimensions with link spins, these are respectively a
surface bounded by a physical loop and a dual surface crossed by the
links to be flipped. Both constructions use the same coordinate sweep.

In a rectangular cubical box, sweep a cell set in one coordinate to the lowest face, recording the next-dimensional cells crossed. The recorded boundary consists of the original set, its projection onto that face, and the pieces swept out by its boundary. For a closed set, the last pieces cancel modulo two. Repeat in the remaining coordinates. Successive projections cancel between sweeps, and the final image at a vertex vanishes in positive dimension. Explicitly, if $C^{(0)}$ is the original closed cell set, $C^{(i)}$ its projection after $i$ sweeps, and $T_i$ the cells recorded in sweep $i$, then $\partial T_i=C^{(i-1)}+C^{(i)}$. Consequently $\partial(T_1+\cdots+T_d)=C^{(0)}+C^{(d)}=C^{(0)}$, because $C^{(d)}$ has no positive-dimensional cells. Thus every closed set of dimension $1,\ldots,d-1$ bounds a filling inside its coordinate box. The argument controls the filling's diameter; it does not require its number of cells to be proportional to the original size.

The same sweep shows that every flat assignment of positive degree on a box is a gauge image. Assign to each $(j-1)$-cell the sum of the given values on the $j$-cells swept out in its full contraction. The sweep boundary identity and flatness imply that the boundary sums reproduce the given assignment. This statement is local to a box; on a torus, winding signs can obstruct a global gauge representation.

Let $c_{d,m},C_{d,m}>0$ be geometric constants depending only on $d,m$, independent of $L,\beta,J$. The first sets a sufficient component-size cutoff, and the second bounds the diameter of a box containing the component. A connected dual $k$-cell component of size $j<c_{d,m}L$ can be lifted consistently to the covering cubic lattice. Choose one cell as the starting cell, or root, and lift along an adjacency spanning tree. An inconsistent extra adjacency would yield a winding path of length $O_{d,m}(j)<L$, impossible for small enough $c_{d,m}$. Its lift fits in a box of diameter $C_{d,m}j$. Shrink $c_{d,m}$ so that $C_{d,m}c_{d,m}<1/2$. Since $k\ge1$, the sweep constructs a filling there. Projecting back gives a nonwrapping local filling.

For each component, fix its local box by a deterministic rule and choose the first valid filling under a fixed ordering of the cell sets inside that box. The correction then depends only on the component and stays local. The analogous statement for a closed primal $m$-cell set uses $m\ge1$. The dual filling itself needs only $1\le k<d$; the stronger $k\le d-2$ in the stability theorem comes from the separate requirement $m\ge1$.

If $a$ and $a'$ produce the same defects, $a\oplus a'$ is flat. Conversely, adding a flat configuration preserves defects. Every nonempty fiber therefore has exactly $|\mathcal A_{\rm flat,L}|$ elements. On a good defect set $E$ with local fillings, let $C(E)$ be the physical set dual to their symmetric difference. Then
\begin{equation}
 a\longmapsto \bigl(E(a),a\oplus C(E(a))\bigr)
 \label{sm:eq:app-fiber-bijection}
\end{equation}
is a bijection between that fiber and all flat configurations. Different fillings can differ by a flat shift, but the chosen deterministic rule fixes the map. The decoder and its reverse use the same rule.

\subsection{A weighted transport statement}
\label{sm:app:weighted}

For a physical cell set $Y$, define its sign character
\begin{equation}
 \sigma_Y(a)\coloneqq\prod_{\ell\in Y}\sigma_\ell=(-1)^{\sum_{\ell\in Y}a_\ell}.
 \label{sm:eq:character}
\end{equation}
Consider a real exponent
\begin{equation}
 F(a)\coloneqq\sum_\alpha k_\alpha\sigma_{Y_\alpha}(a),
 \qquad k_\alpha\in\mathbb R.
 \label{sm:eq:weighted-F}
\end{equation}
Subscripted $k$'s such as $k_\alpha$ denote expansion coefficients; unindexed $k$ remains the defect dimension. A scalar term is separated because it cancels from the probability law. Each nonempty character has an integer generating cost $\nu_\alpha\ge1$ and an assigned connected carrier $K_\alpha$ containing $Y_\alpha$, with
$|K_\alpha|\le b_w\nu_\alpha$ and
$\diam K_\alpha\le d_w\nu_\alpha$.
A carrier includes the support and fixed local paths of incident lattice cells that connect it; the paths may contain cells of different dimensions and add no physical spins. Include the boundary $(m-1)$-cells of its physical cells as well. This changes only fixed geometric constants and ensures that carriers sharing a gauge charge actually meet. Distinct generating labels are retained even when they give the same final character. For a fixed exponential reserve $S$, define
\begin{equation}
 Q_S\coloneqq\sup_{L,\ell}
       \sum_{\alpha:\ell\in K_\alpha}|k_\alpha|e^{S\nu_\alpha}.
 \label{sm:eq:weighted-norm}
\end{equation}

For the quantum application, an enlarged support consists of the original support of a perturbing term together with all physical cells in reference interactions $B_p$ meeting that support. It contains the term's support after imaginary-time conjugation by $H_0$ and has size and diameter bounded independently of $L$. This enlarged support need not be connected; adding fixed local paths gives a connected carrier. When the enlarged supports form a connected overlap graph, the union of their carriers obeys the linear bounds above, after adjusting $b_w,d_w$.

Enlarging finite-range supports to connected carriers of bounded size changes only class-dependent constants. A diagonal term on at most $b$ qubits has at most $2^b$ sign characters, each coefficient bounded by its supremum norm. Thus $F=-\beta gV$ satisfies $Q_S\le C_S\beta|g|$ with cost one after adjusting $b_w,d_w$.

\begin{proposition}[Weighted classical equivalence]
\label{sm:prop:weighted}
For fixed $d,m,b_w,d_w$ with $1\le m\le d-2$, there are constants $B_w,a_w>0$ and a finite reserve $S$ such that
\begin{equation}
 \beta J>B_w,\qquad Q_S<a_w
 \label{sm:eq:weighted-window}
\end{equation}
imply $\boldsymbol p_F\sim_{\LC}\boldsymbol\rho_{\rm flat}$, where $p_F$ is defined in Eq.~\eqref{sm:eq:classical-density}. For every fixed $s>0$, both conversion errors are $O(L^{-s})$, with polylogarithmic block diameter, depth, and causal radius and product work-register inputs.
\end{proposition}

The retained generating labels, connected carriers, and exponential cost reserve are shared with the weighted classical expansion in Ref.~\cite{SM-WatanabeThermodynamics2026}. That expansion controls the bulk free energy; Proposition~\ref{sm:prop:weighted} additionally requires local conversion maps and their whole-system accuracy bounds.

We prove this proposition in the following four subsections. The constants are chosen successively to ensure convergence of gauge averaging, rapid orbit sampling, a weak interaction in the positive defect law, and contraction of the defect sampler. In particular, the explicit unperturbed Peierls threshold alone is not asserted to suffice for this two-way statement.

\subsection{Gauge averaging and the conditional inverse}
\label{sm:app:orbit}

The conditional gauge law recovers the state after gauge averaging; we prove a local sampling estimate uniform in the physical input. The Boltzmann-weight average~\cite{SM-FoersterNielsenNinomiya1980,SM-Thorngren2026} is shared with Ref.~\cite{SM-WatanabeThermodynamics2026}.

Use the gauge bits $\widehat\eta_u\coloneqq(1-\eta_u)/2\in\{0,1\}$ associated with the signs $\eta_u$ in Eq.~\eqref{sm:eq:gauge-action}, including all redundant parameters on the $(m-1)$-cells $u$. A physical cell $\ell$ is flipped by the parity $\sum_{u\in\partial\ell}\widehat\eta_u$; the resulting bit configuration is $a\oplus\delta\widehat\eta$. The action preserves every $B_p$ and hence $H_0$, since $\delta^2\widehat\eta=0$. No gauge invariance of $F$ or $V$ is required. Each parameter touches $2(d-m+1)$ physical cells, and a fixed coloring implements all independent gauge flips in bounded depth.

The exact averaged perturbation weight and the conditional gauge law are
\begin{align}
 w_F(a)&\coloneqq2^{-n_{m-1}}\sum_{\widehat\eta}
          e^{F(a\oplus\delta\widehat\eta)},\label{sm:eq:gauge-weight}\\
 (\mathcal Gp_F)(a)&\propto e^{\beta J\sum_pB_p(a)}w_F(a),\label{sm:eq:gauge-average}\\
 Q_a(\widehat\eta)&\coloneqq
 \frac{e^{F(a\oplus\delta\widehat\eta)}}{2^{n_{m-1}}w_F(a)}.
 \label{sm:eq:orbit-conditional}
\end{align}
Here $\mathcal G$ samples the gauge bits independently and uniformly and applies their local flips.

For $a$ drawn from $\mathcal Gp_F$, sampling $Q_a$ and applying its gauge flip restores $p_F$ exactly. Indeed, for a proposed output $a'$, put $a=a'\oplus\delta\widehat\eta$. Substituting into the joint density of $a,\widehat\eta$ cancels $w_F(a)$, leaving a factor proportional to $e^{\beta J\sum_pB_p(a')+F(a')}$ independent of $\widehat\eta$. Summing all $2^{n_{m-1}}$ choices restores $p_F(a')$. No injectivity of $\widehat\eta\mapsto\delta\widehat\eta$ is used. Both laws still range over every physical spin configuration; gauge averaging introduces neither a quotient nor a Gauss-law constraint.

To bound conditional sampling influences, express each character in the gauge bits:
\begin{equation}
 \sigma_{Y_\alpha}(a\oplus\delta\widehat\eta)
 =\sigma_{Y_\alpha}(a)\prod_{u\in\partial Y_\alpha}\eta_u.
 \label{sm:eq:gauge-character}
\end{equation}
Here $\prod_{u\in\partial Y_\alpha}\eta_u=(-1)^{\sum_{u\in\partial Y_\alpha}\widehat\eta_u}$. The gauge charge set $\partial Y_\alpha$ has at most $2mb_w\nu_\alpha$ cells. For fixed physical input $a$, write the single-site conditional law as $q_{a,u}(s\mid\xi)\coloneqq Q_a(\widehat\eta_u=s\mid\widehat\eta_{\ne u}=\xi)$ and define the dimensionless Dobrushin influence coefficient
\begin{equation}
 c_{uu'}\coloneqq\max_{\substack{\xi,\xi':\ \xi_v=\xi'_v\\v\ne u,u'}}
 \bigl\|q_{a,u}(\cdot\mid\xi)-q_{a,u}(\cdot\mid\xi')\bigr\|_{\rm TV}
 \quad(u'\ne u).
 \label{sm:eq:dobrushin-coefficient}
\end{equation}
The maximum ranges over assignments off $u$ differing only at $u'$, and
$\|p-q\|_{\rm TV}\coloneqq\tfrac12\sum_{s=0}^1|p(s)-q(s)|$; the dependence of $c_{uu'}$ on $a$ is suppressed. In the binary single-site conditional law, this coefficient is at most the sum of $\tanh|k_\alpha|$ over characters containing both variables. Summing the row and using the fixed number of physical cofaces yields
\begin{equation}
 \sup_u\sum_{u'\ne u}c_{uu'}\le C Q_S<\tfrac12
 \label{sm:eq:orbit-influence}
\end{equation}
for small enough $Q_S$. This Dobrushin estimate~\cite{SM-Dobrushin1968} is uniform in $a$, because $\sigma_{Y_\alpha}(a)$ changes only the coefficient's sign.

Cut off labels with cost larger than
$r_{\rm cut}\coloneqq C\log(n/\varepsilon)$. Their total exponent norm is at most $CnQ_Se^{-cr_{\rm cut}}$. A uniform exponent change of size $u$ changes normalized probabilities in $\ell^1$ by at most $e^{2u}-1$, so this cutoff costs $O(\varepsilon)$. Remaining updates have range $O(r_{\rm cut})$.

Color intersecting update neighborhoods with $O(r_{\rm cut}^d)$ colors and use independent random coins for every update. A maximal coupling of each conditional update contracts the maximum disagreement probability by a fixed factor after a full sweep: during a sweep it never exceeds its initial maximum, and every updated site is at most the row bound times that maximum. After $C\log(n/\varepsilon)$ sweeps, a union bound over gauge variables gives total variation error $O(\varepsilon)$. This proves a local approximate inverse of the gauge average uniformly for every input $a$.

\subsection{The connected expansion and its flux potential}
\label{sm:app:flux-potential}

We sum the perturbing clusters into a real flux potential and prove the local norm in Eq.~\eqref{sm:eq:weighted-flux-potential}, which controls the positive defect sampler and its conversion error. The neutral-character expansion and local-filling identity are shared with Ref.~\cite{SM-WatanabeThermodynamics2026}, where they enter an exact component sum for the partition function.

The binary factorization is exact:
\begin{equation}
 e^{k_\alpha\sigma_{Y_\alpha}}
 =\cosh k_\alpha\bigl(1+t_\alpha\sigma_{Y_\alpha}\bigr),
 \qquad t_\alpha\coloneqq\tanh k_\alpha.
 \label{sm:eq:binary-factor}
\end{equation}
Expand the product and average the gauge variables. Each independent gauge sign contributes zero if it occurs an odd number of times and one otherwise. Thus a selected set of labels survives precisely when the symmetric difference of its charge sets $\partial Y_\alpha$ is empty. Join selected labels whose nonempty charge sets meet; charge-free labels form singleton components. Each connected component is neutral on its own. Call such a component a polymer. Two polymers are incompatible if they share an original label or a charged gauge cell. Its activity is the product of its $t_\alpha$ factors times its physical character. This gives an exact hard-core polymer representation of $w_F/\prod_\alpha\cosh k_\alpha$.

The convergence estimate depends on the generating labels rather than only on the final sign product. For a label of cost $\nu$, the weighted sum of labels whose charges can meet it is at most $C\nu Q_S$. The same bound covers reuse of a label. Fix an exponential weight $e^{s_0\nu}$ to retain in the connected logarithm, with $s_0>0$, and choose $S\ge s_0+2$. Starting from zero, iterate the rooted-tree recursion
\begin{equation}
 T_\alpha=|t_\alpha|e^{(1+s_0)\nu_\alpha}\exp\!\left(
                         \sum_{\alpha'\sim\alpha}T_{\alpha'}\right).
 \label{sm:eq:tree-recursion}
\end{equation}
Here $\alpha'\sim\alpha$ means identical labels or intersecting charge sets. The iterates are bounded by $|t_\alpha|e^{(2+s_0)\nu_\alpha}$: substituting this bound in the neighboring sum gives at most $CQ_S\nu_\alpha\le\nu_\alpha$, since $2+s_0\le S$ and $Q_S$ is sufficiently small. Their monotone limit is the tree sum. The sum over neutral connected sets is bounded by this tree sum. Summing absolute activities of polymers incompatible with a fixed polymer, weighted by the exponential of $(1+s_0)$ times their total cost, is consequently bounded by $CQ_S$ times the fixed polymer's cost. This is the Koteck\'y--Preiss criterion~\cite{SM-KoteckyPreiss1986} with the reserve $s_0$ included, giving an absolutely convergent connected logarithm.

Explicitly, a logarithmic cluster is a list of polymers whose incompatibility graph is connected, with coefficient given by the sum over connected spanning subgraphs of the product of $-1$ on incompatible edges, divided by the list factorial. Repeated polymers in this logarithm are permitted and their costs are counted repeatedly. Marking a polymer occurrence, applying the convergence bound with that occurrence held fixed, and then using the label-root tree bound gives the local estimate
\begin{equation}
 \sup_\ell\sum_{\Gamma:\ell\in K_\Gamma}
       \norm{\phi_\Gamma}_\infty e^{s_0\nu_\Gamma}\le CQ_S.
 \label{sm:eq:classical-pinned-log}
\end{equation}
Here $\phi_\Gamma(a)$ is the local contribution to the connected logarithm, including the connected-graph coefficient, the perturbing factors, and their physical sign character. The supremum norm is over all spin configurations $a$, $\nu_\Gamma$ is total generating cost, and $K_\Gamma$ is the union of generating carriers. The reserve $s_0$ is the one chosen before Eq.~\eqref{sm:eq:tree-recursion}. This local bound, rather than only an extensive partition-function bound, permits the root relocation used below.

Every resulting term has a closed physical $m$-cell character, because each polymer is charge neutral. If its total generating cost is $j$, its assigned connected carrier has diameter $O(j)$. For $j<cL$, a sufficiently small fixed $c$ makes that carrier nonwrapping. The local filling statement then supplies an $(m+1)$-cell set $S_Y$ with $\partial S_Y=Y$, and cancellation of interior cells gives
\begin{equation}
 \sigma_Y(a)=\prod_{p\in S_Y}B_p(a)
           =(-1)^{|S_Y\cap E^*(a)|}.
 \label{sm:eq:local-Stokes}
\end{equation}
Here $E^*(a)$ denotes the primal interaction cells dual to $E(a)$. This equation also defines the term on defect assignments not satisfying the physical constraints. Only one local filling is chosen per term; fillings are not enumerated.

Separating scalar terms and applying Eq.~\eqref{sm:eq:local-Stokes} to nonwrapping terms gives a local flux potential; the remaining large-cost terms form $r_L$:
\begin{align}
 \log w_F(a)&=c_L-U(E(a))+r_L(a),\\
 U(E)&\coloneqq\sum_X\Psi_X(E|_X),\\
 \norm{r_L}_\infty&\le CnQ_Se^{-cL},\\
 M_\mu&\coloneqq\sup_{p^\ast}\sum_{X\ni p^\ast}|X|e^{\mu\diam X}
                                  \norm{\Psi_X}_\infty\le CQ_S.
 \label{sm:eq:weighted-flux-potential}
\end{align}
Here $c_L$ is configuration independent; $X$ ranges over sets of dual defect cells, $E|_X$ records their occupations, and $\Psi_X$ is a real function of these occupations, defined for arbitrary assignments. The positive constant $\mu$ is chosen sufficiently small and independently of $L$.

To check the last bound, choose $C_0$ so that the filling box of a cost-$j$ term has diameter at most $C_0j$ and hence contains at most $Cj^d$ defect cells. Relocating its root to a specified defect cell costs another factor at most $Cj^d$. Including $|X|$ and $e^{\mu\diam X}$ is therefore paid for by the convergent bound
$CQ_S\sum_jj^{2d}e^{-(s_0-\mu C_0)j}$, with fixed $s_0>\mu C_0$. This is why the generating carriers and exponential reserve must be retained even if a final character has very small support.

\subsection{A positive defect law, cutoff, and decoder}
\label{sm:app:positive-gas}

To estimate component probabilities by deletion, temporarily allow all closed dual $k$-cell sets, including ones that are not boundaries. This distinction is needed on a torus: two winding components can together be a boundary although either alone is not. Deleting one can therefore violate the physical constraint, even though it preserves closure. The weighted potential is defined on all defect assignments, so the enlarged law is well defined:
\begin{equation}
 \pi(E)\coloneqq
 \frac{\one_{\{\partial E=\varnothing\}}e^{-2\beta J|E|-U(E)}}
      {\sum_{E':\,\partial E'=\varnothing}e^{-2\beta J|E'|-U(E')}}.
 \label{sm:eq:aux-law}
\end{equation}
Relative to the physical defect law with potential $U$, only the global boundary constraint has been relaxed; closure and the weights are unchanged.

Fix a closed connected set $\gamma$ and consider configurations in which it is a connected component. The map $E\mapsto E\setminus\gamma$ preserves closure and is injective, since adding the fixed $\gamma$ back recovers $E$. Only potentials meeting $\gamma$ can change, and Eq.~\eqref{sm:eq:weighted-flux-potential} gives
\begin{equation}
 |U(E)-U(E\setminus\gamma)|
 \le2\sum_{X:\,X\cap\gamma\ne\varnothing}\norm{\Psi_X}_\infty
 \le2M_\mu|\gamma|.
 \label{sm:eq:component-potential-change}
\end{equation}
Thus the input weight is at most $e^{-(2\beta J-2M_\mu)|\gamma|}$ times the weight after deletion. Summing the latter weights over the deletion image is bounded by the full denominator in Eq.~\eqref{sm:eq:aux-law}. Put
\begin{equation}
 \kappa_0\coloneqq2\beta J-2M_\mu,
 \qquad A_j\coloneqq D_{d,m}^{2(j-1)}.
 \label{sm:eq:gas-parameters}
\end{equation}
The deletion bound and the connected-set count in Sec.~\ref{sm:sec:model} therefore give
\begin{align}
 \Pr_\pi(\gamma\text{ is a component})&\le e^{-\kappa_0|\gamma|},\\
 \Pr_\pi(\exists\gamma:|\gamma|\ge r_{\rm cut})
 &\le n_{m+1}\sum_{j\ge r_{\rm cut}}A_je^{-\kappa_0j}.
 \label{sm:eq:weighted-tail}
\end{align}
Choose the low-temperature, weak-interaction domain so that $\kappa_0>2\log D_{d,m}$. If every component has fewer than $c_{d,m}L$ cells, the sum modulo two of its local fillings fills $E$. Hence a closed set that is not a boundary must contain a component of size at least $\lceil c_{d,m}L\rceil$. Its probability is exponentially small by Eq.~\eqref{sm:eq:weighted-tail}. Conditioning on the boundary constraint changes the normalized law by exactly
\begin{equation}
 \norm{\pi-\pi(\,\cdot\mid E\text{ is a boundary})}_1
 =2\pi(E\text{ is not a boundary})=O(ne^{-cL}),
 \label{sm:eq:conditioning-error}
\end{equation}
where $c>0$ is independent of $L$ in the chosen domain. Dropping $r_L$ in Eq.~\eqref{sm:eq:weighted-flux-potential} has the same order of error. Equal fiber cardinalities then identify the conditioned law, lifted uniformly to each defect fiber, with the gauge-averaged density up to this error.

For local implementation, truncate potentials at diameter $r\coloneqq C\log(n/\varepsilon)$, calling the result $U_r$ and its closed-set law $\pi_r$. Define $r_{\rm cut}\coloneqq\lceil C\log(n/\varepsilon)\rceil$ and let $G_{r_{\rm cut}}$ be the event that every component has size at most $r_{\rm cut}$. The cutoff law is the conditional distribution $\pi_{\rm cut}\coloneqq\pi_r(\,\cdot\mid G_{r_{\rm cut}})$. It is not obtained by erasing large components from a sampled configuration.

Potential deletion changes the total exponent by at most $CnM_\mu e^{-\mu r}$, and Eq.~\eqref{sm:eq:weighted-tail} also applies to $U_r$. Thus $\norm{\pi_r-\pi_{\rm cut}}_1=2\pi_r(G_{r_{\rm cut}}^c)$, and large enough cutoff constants make both errors $O(\varepsilon)$. For $\varepsilon=L^{-s}$ with fixed $s>0$, the cutoff satisfies $r_{\rm cut}<c_{d,m}L$ for sufficiently large $L$. All retained components then have local fillings and satisfy the physical constraint automatically.

Lift $\pi_{\rm cut}$ uniformly onto each defect fiber. Under Eq.~\eqref{sm:eq:app-fiber-bijection}, this lifted law is exactly
\begin{equation}
 \pi_{\rm cut}\otimes\mathrm{Uniform}(\mathcal A_{\rm flat,L}).
 \label{sm:eq:cut-factorization}
\end{equation}
The decoder inspects neighbors in an unchanged copy of the original defect set. It stops an exploration when all neighbors of discovered cells have been checked or $r_{\rm cut}+1$ occupied cells have been found. Only an exhausted exploration with at most $r_{\rm cut}$ cells is corrected; this identifies a connected component rather than a fragment of a larger one. Apply each component's correction once, using its first cell in a fixed ordering as its representative. It requires radius $O(r_{\rm cut})$ per output and $O(r_{\rm cut}^d)$ colors. It maps the cutoff law exactly to $\rho_{\rm flat,L}$, so contractivity bounds the actual whole-state error by $O(\varepsilon)$. This proves the forward conversion.

\subsection{Reverse sampling, contraction, and finite circuits}
\label{sm:app:reverse-sampling}
\label{sm:app:reverse-sampler}

The reverse channel starts with the flat input and an independent sampler of $\pi_{\rm cut}$. We use component birth and death, as in loss-network samplers~\cite{SM-FernandezFerrariGarcia2002}; the contraction and finite-circuit bounds needed here are proved below. Distinct components are compatible when they neither overlap nor share a $(k-1)$-face; their union then has exactly those connected components. On collections of compatible closed components of size at most $r_{\rm cut}$, remove each present component at rate one and insert an absent compatible $\gamma$ at rate
\begin{equation}
 b_\gamma(E)\coloneqq
 e^{-2\beta J|\gamma|-U_r(E\cup\gamma)+U_r(E)}
 \le e^{-\kappa_0|\gamma|}.
 \label{sm:eq:app-birth-rate}
\end{equation}
The ratio is exactly that of the target weights, proving detailed balance. Deletions and insertions connect every state to the empty collection, so the stationary law is uniquely $\pi_{\rm cut}$.

To prove rapid mixing, use the sum of sizes of unmatched components as the distance. Compare collections differing by one component $\gamma'$, and couple common deaths and births. The unmatched death has drift $-|\gamma'|$. For incompatible births, select a touching cell and re-root the connected-set count there. There are at most $(D_{d,m}+1)|\gamma'|$ possible touching cells. Define
\begin{equation}
 S_{\kappa_0}\coloneqq\sum_{j\ge1}jA_je^{-\kappa_0j}.
 \label{sm:eq:sampler-series}
\end{equation}
Their positive drift is at most
$(D_{d,m}+1)S_{\kappa_0}|\gamma'|$.

For a birth $\gamma$ compatible with both $E$ and $E\cup\gamma'$, the difference of its log rates is
\begin{equation}
 \log b_\gamma(E\cup\gamma')-\log b_\gamma(E)
 =-U_r(E\cup\gamma\cup\gamma')+U_r(E\cup\gamma')
   +U_r(E\cup\gamma)-U_r(E).
 \label{sm:eq:birth-mixed-difference}
\end{equation}
A potential cancels from this four-term difference unless it meets both components; otherwise its contribution is at most $4\norm{\Psi_X}_\infty$.
Using $|e^{-u}-e^{-v}|\le\max(e^{-u},e^{-v})|u-v|$ and the birth-rate bound gives
\begin{equation}
 |b_\gamma(E\cup\gamma')-b_\gamma(E)|
 \le4e^{-\kappa_0|\gamma|}
 \sum_{\substack{X:\,X\cap\gamma\ne\varnothing\\X\cap\gamma'\ne\varnothing}}
                  \norm{\Psi_X}_\infty.
 \label{sm:eq:birth-rate-influence}
\end{equation}
Multiply by $|\gamma|$ to bound the positive distance drift and sum over candidate births. For each fixed defect cell $y$, the connected-set count gives
$\sum_{\gamma\ni y}|\gamma|e^{-\kappa_0|\gamma|}\le S_{\kappa_0}$.
Choose a cell of $\gamma'$ and then a potential containing it; the factor $|X|$ in Eq.~\eqref{sm:eq:weighted-flux-potential} pays for the choice of $y\in X$. Consequently,
\begin{equation}
 \sum_{\gamma\text{ compatible with both}}|\gamma|
       |b_\gamma(E\cup\gamma')-b_\gamma(E)|
 \le4M_\mu S_{\kappa_0}|\gamma'|.
 \label{sm:eq:birth-drift-bound}
\end{equation}
 Choose the constants so
\begin{equation}
 (D_{d,m}+1+4M_\mu)S_{\kappa_0}<\tfrac12.
 \label{sm:eq:sampler-contraction}
\end{equation}
Path coupling~\cite{SM-BubleyDyer1997} then yields a fixed contraction rate. Deleting unmatched components and adding target components gives a path within the compatible state space, so the adjacent-pair estimate applies globally. The distance is at least one off the diagonal and at most $2n_{m+1}$. Starting from empty flux, total variation distance to stationarity is at most $2n_{m+1}e^{-ct}$. Hence
$t\coloneqq C\log(n/\varepsilon)$ suffices.

We now turn the continuous-time sampler into a finite local circuit. Assign each candidate component to its first cell in a fixed ordering; this cell is its root. At each root, an independent marked Poisson process proposes births with rates $e^{-\kappa_0|\gamma|}$; its total rate is bounded by $\sum_jA_je^{-\kappa_0j}$. Reject incompatible proposals and accept a compatible proposal with probability $b_\gamma(E)e^{\kappa_0|\gamma|}\le1$, implementing Eq.~\eqref{sm:eq:app-birth-rate}. A separate rate-one clock deletes the component currently rooted there, if one exists. At most one present component has a given root. Every event depends on records within radius
$r_u\coloneqq O(r+r_{\rm cut})$.

For a chronological dependency path of $n_{\rm ev}$ events, where $n_{\rm ev}$ is a positive integer, each next root has at most $Cr_u^d$ choices, and the ordered event times have volume $t^{n_{\rm ev}}/n_{\rm ev}!$. Consequently
\begin{equation}
 \Pr(\text{some path has }n_{\rm ev}\text{ events})
 \le Cn\frac{(Cr_u^dt)^{n_{\rm ev}}}{n_{\rm ev}!}.
 \label{sm:eq:dependency-bound}
\end{equation}
Choose $n_{\rm ev}^*\coloneqq C[r_u^dt+\log(n/\varepsilon)]$. Stirling's bound makes the probability of a path longer than $n_{\rm ev}^*$ of order $\varepsilon$. Poisson tails also permit a polylogarithmic cap on each local record length.

Discretize event times on a mesh $\delta_t\le\varepsilon^2/[Cn^2(t+1)]$. The expected number of pairs sharing a time bin is at most $Cn^2(t+1)\delta_t$, so even counting all roots this changes the chronological process with probability $O(\varepsilon)$. Approximate every proposal mark distribution and acceptance coin in total variation to accuracy at most $\varepsilon/[Cn(t+1)]$. The expected total number of proposals is $O(n(t+1))$, which bounds their combined error by $O(\varepsilon)$. The allowed block maps need not implement these finite records with a fixed gate alphabet.

Store all records in independent local work registers. In each of $n_{\rm ev}^*$ rounds, use neighborhood colors to process at most the earliest ready event at each root, where readiness means that all its earlier overlapping dependencies have been processed. Events at the same root are ordered too, so the dependency-height cutoff ensures that $n_{\rm ev}^*$ rounds suffice on the good records. All gates act on polylogarithmic blocks; the number of rounds and colors is polylogarithmic. If a local record cannot be completed within the fixed budget, a predetermined local fallback output is used. No global success flag is required and there is no postselection. Coupling to the original process on the good records proves total variation error $O(\varepsilon)$.

Finally add the sampled fillings to the flat input, discard the sampling registers, and apply the approximate conditional inverse of gauge averaging. Equation~\eqref{sm:eq:cut-factorization} gives the intended joint law before the inverse.

Assign separate $O(\varepsilon)$ budgets to the wrapping remainder, physical-sector conditioning, potential and component cutoffs, mixing, event discretization, and orbit sampling. Subsequent channels contract trace norm, so these errors add. Taking $\varepsilon=L^{-s}$ proves Proposition~\ref{sm:prop:weighted} and the classical part of Theorem~\ref{sm:thm:main}.
\section{Diagonal reduction and deterministic coherence recovery}
\label{sm:app:quantum}

We prove the quantum extension in two steps. First the diagonal probabilities acquire a weak classical interaction satisfying Proposition~\ref{sm:prop:weighted}. Then an auxiliary gapped path supplies a physical local channel recovering the off-diagonal matrix elements. Gap stability and spectral flow are external theorems; their required estimates and the distinction between physical and auxiliary Hamiltonians are made explicit below.

\subsection{A connected expansion for diagonal probabilities}

This diagonal reduction shares its Duhamel histories, enlarged supports, connected logarithm, and generating-cost bounds with the quantum proof in Ref.~\cite{SM-WatanabeThermodynamics2026}. There the resulting classical interaction is used to prove analyticity of the bulk free energy. Here it supplies the diagonal state for Proposition~\ref{sm:prop:weighted}; the auxiliary parents and controlled spectral flow in the following subsections provide the additional deterministic recovery of quantum coherences.

Let $\mathcal D_{\rm diag}$ be onsite $Z$-basis dephasing and define
\begin{align}
 \tau_g&\coloneqq\mathcal D_{\rm diag}(\rho_{\beta,g,L}),\\
 \mathcal W_{\beta,g}(a)&\coloneqq e^{\beta H_0(a)}\bra a e^{-\beta H_g}\ket a.
 \label{sm:eq:quantum-diagonal}
\end{align}
The diagonal law is proportional to $e^{-\beta H_0(a)+\log\mathcal W_{\beta,g}(a)}$. Its interaction-picture terms and small parameter are
\begin{equation}
 v_X(u)\coloneqq e^{u\beta H_0}v_Xe^{-u\beta H_0},\qquad 0\le u\le1,
 \label{sm:eq:dressed-term}
\end{equation}
\begin{equation}
 \norm{v_X(u)}\le e^{4b(d-m)\beta J},\qquad
 \lambda\coloneqq\beta|g|e^{4b(d-m)\beta J}.
 \label{sm:eq:lambda}
\end{equation}

For the positive ratio $\mathcal W_{\beta,g}(a)$ in Eq.~\eqref{sm:eq:quantum-diagonal}, Duhamel expansion gives
\begin{equation}
 \mathcal W_{\beta,g}(a)=\bra a\mathcal T\exp\!\left[
   -\beta g\int_0^1du\sum_Xv_X(u)\right]\ket a,
 \label{sm:eq:duhamel}
\end{equation}
where $\mathcal T$ puts larger $u$ to the left and $v_X(u)$ is defined in Eq.~\eqref{sm:eq:dressed-term}. To obtain the identity, differentiate $e^{u\beta H_0}e^{-u\beta H_g}$: its derivative is obtained by multiplying it from the left by $-\beta g\sum_Xv_X(u)$, and its initial value is $I$. Iteration gives the time-ordered exponential.

The enlarged support described in Sec.~\ref{sm:app:weighted} is the fixed set
\begin{equation}
 U_X\coloneqq X\cup\bigcup_{p:\partial p\cap X\ne\varnothing}\partial p.
 \label{sm:eq:enlarged-support}
\end{equation}
Only the $B_p$ with $\partial p\cap X\ne\varnothing$ enter the conjugation, so $\supp v_X(u)\subset U_X$. Each such interaction contains at most $2m+1$ physical cells beyond a chosen cell of $X$. Thus $U_X$ contains at most
\begin{equation}
 b_h\coloneqq b[1+2(d-m)(2m+1)]
 \label{sm:eq:enlarged-support-size}
\end{equation}
physical cells and has fixed diameter. At most $2b(d-m)$ reference interactions enter the conjugation. Each contributes at most $e^{2\beta J}$ to the norm bound, giving Eq.~\eqref{sm:eq:lambda}.

Form a graph whose vertices are the interaction labels $X$, joining two labels when their enlarged supports overlap. This graph has bounded degree, and only a fixed number of enlarged supports contain a given physical cell. Add fixed local paths to each $U_X$ to obtain a connected carrier $K_X$. These carriers have bounded size $b_c$, diameter, and incidence, depending only on the interaction class.

For each nonempty connected label set $\mathcal A$, define $w_{\mathcal A}(a)$ as the sum of terms in Eq.~\eqref{sm:eq:duhamel} whose distinct interaction labels are exactly $\mathcal A$. A label may occur repeatedly. With fixed multiplicities $n_X\ge1$, the ordered simplex has volume $1/(\sum_X n_X)!$ and there are $(\sum_X n_X)!/\prod_X n_X!$ label orderings. Their combined absolute weight is therefore at most $\prod_{X\in\mathcal A}\lambda^{n_X}/n_X!$. Summing each $n_X$ from one to infinity gives
\begin{equation}
 |w_{\mathcal A}(a)|\le(e^\lambda-1)^{|\mathcal A|}.
 \label{sm:eq:quantum-activity}
\end{equation}

Histories on disjoint enlarged supports factor in the product vector $\ket a$, including the shuffles of their time orderings. Thus $\mathcal W_{\beta,g}$ is exactly a hard-core polymer partition function, with incompatibility given by overlap of the unions $\bigcup_{X\in\mathcal A}U_X$. The activities are real: inclusion--exclusion over subsets of a connected label set expresses an activity as a real linear combination of diagonal matrix elements of Gibbs operators for Hermitian partial Hamiltonians, times the real reference factor. This remains true when the perturbation matrices are complex.

To bound the connected logarithm, let $D_h$ bound the degree of this overlap graph and $c_h^{j-1}$ bound the number of connected $j$-label sets containing one label. For a fixed reserve $A>0$, put
\begin{equation}
 t_A\coloneqq(e^\lambda-1)e^{A+1}.
 \label{sm:eq:quantum-tA}
\end{equation}
The weighted convergence criterion is satisfied if
\begin{equation}
 (D_h+1)\frac{t_A}{1-c_ht_A}\le1,
 \qquad c_ht_A<1.
 \label{sm:eq:quantum-KP}
\end{equation}
Indeed the sum of weighted activities incompatible with $\mathcal A$ is at most the left side times $|\mathcal A|$. The connected logarithm expansion~\cite{SM-KoteckyPreiss1986} therefore gives
\begin{align}
 \log \mathcal W_{\beta,g}(a)&=\sum_\Gamma\phi_\Gamma(a),
 \label{sm:eq:quantum-log}\\
 \sup_{L,\ell}\sum_{\Gamma:\ell\in K_\Gamma}
 e^{A\nu_\Gamma}\norm{\phi_\Gamma}_\infty
 &\le C_A\lambda.
 \label{sm:eq:quantum-weighted-cluster}
\end{align}
Here $\Gamma$ is a connected logarithmic cluster of polymers,
$\nu_\Gamma\coloneqq\sum_{\mathcal A\in\Gamma}|\mathcal A|$ counts repeated labels between its polymers, and $Y_\Gamma\coloneqq\bigcup_{\mathcal A\in\Gamma}\bigcup_{X\in\mathcal A}U_X$ is the union of their enlarged supports. Define $K_\Gamma\coloneqq\bigcup_{\mathcal A\in\Gamma}\bigcup_{X\in\mathcal A}K_X$, with each occurrence of a label included when computing the cost. This union is connected because the enlarged supports in a logarithmic cluster overlap along its connected graph. We have $|Y_\Gamma|\le b_h\nu_\Gamma$, $|K_\Gamma|\le b_c\nu_\Gamma$, and $\diam K_\Gamma=O(\nu_\Gamma)$.

The bound in Eq.~\eqref{sm:eq:quantum-weighted-cluster} follows by summing activities of polymers whose carrier contains the fixed physical cell. A fixed number of original carriers contain that cell, so the convergence bound with one label held fixed gives a fixed incidence factor times $t_A/(1-c_ht_A)$. The positive ratio $\mathcal W_{\beta,g}$ fixes the real logarithm continuously from $g=0$.

Convert this local logarithm to the sign-character form required by Proposition~\ref{sm:prop:weighted}:
\begin{align}
 \phi_\Gamma(a)&=\sum_{Y\subset Y_\Gamma}k_{\Gamma,Y}\sigma_Y(a),\\
 \sum_{Y\subset Y_\Gamma}|k_{\Gamma,Y}|
 &\le2^{|Y_\Gamma|}\norm{\phi_\Gamma}_\infty.
 \label{sm:eq:quantum-Fourier}
\end{align}
Retain a separate label with carrier $K_\Gamma$ and its original generating cost for every coefficient, even if cancellation makes $Y$ much smaller than $Y_\Gamma$. After separating scalar terms and choosing $A>S+b_h\log2$, Eq.~\eqref{sm:eq:quantum-weighted-cluster} gives the norm $Q_S\le C_S\lambda$ of Eq.~\eqref{sm:eq:weighted-norm}. Proposition~\ref{sm:prop:weighted} consequently applies to the exact diagonal density $\tau_g$.

\newpage
\subsection{Local auxiliary parents and a uniform gap}

Turn on the perturbation with dimensionless path time $0\le t\le1$, keeping $\beta$ fixed, and define
\begin{align}
 H(t)&\coloneqq H_0+tgV,\label{sm:eq:conditional-path}\\
 M_t&\coloneqq e^{-\beta H(t)/2},\\
 d_t(a)&\coloneqq\bra a M_t^2\ket a,\\
 \ket{\psi_{t,a}}&\coloneqq\frac{M_t\ket a}{\sqrt{d_t(a)}},\\
 p_t(a)&\coloneqq\frac{d_t(a)}{\Tr M_t^2}.
 \label{sm:eq:conditional-vectors}
\end{align}
At $t=0$, $\ket{\psi_{0,a}}=\ket a$; at $t=1$, $\tau_g=\sum_a p_1(a)\proj a$. Moreover, $p_t(a)\proj{\psi_{t,a}}=M_t\proj a M_t/\Tr M_t^2$. Summing over $a$ gives the product-basis thermal-ensemble identity~\cite{SM-White2009}
\begin{equation}
 \frac{M_t^2}{\Tr M_t^2}
 =\sum_a p_t(a)\proj{\psi_{t,a}}.
 \label{sm:eq:conditional-ensemble}
\end{equation}

The auxiliary conditional parents are
\begin{align}
 P_{\ell}(a)&\coloneqq I-\proj{a_{\ell}},\\
 Q_{\ell}(t,a)&\coloneqq M_tP_{\ell}(a)M_t^{-1},\\
 \mathcal K_a(t)&\coloneqq\sum_{\ell}Q_{\ell}(t,a)^\dagger Q_{\ell}(t,a).
 \label{sm:eq:auxiliary-parent}
\end{align}

The parent has exactly one zero-energy state. Positivity of every $Q_\ell^\dagger Q_\ell$ shows that the kernel is the common zero space of all $Q_\ell$. Here $P_\ell(a)$ rejects the specified bit at site $\ell$. Since $M_t$ is invertible, $Q_\ell\ket\phi=0$ for every $\ell$ is equivalent to $P_\ell M_t^{-1}\ket\phi=0$ for every $\ell$. Thus $M_t^{-1}\ket\phi$ must be proportional to $\ket a$, and
\begin{equation}
 \ker\mathcal K_a(t)=\operatorname{span}\{M_t\ket a\},
 \qquad \mathcal K_a(0)=\sum_\ell P_\ell(a).
 \label{sm:eq:parent-kernel}
\end{equation}
This proves uniqueness but not gap stability.

To prove locality and smallness, define
\begin{equation}
 T_t(u)\coloneqq e^{-u\beta H(t)/2}e^{u\beta H_0/2},
 \qquad 0\le u\le1.
 \label{sm:eq:intertwiner}
\end{equation}
Here $u$ is dimensionless imaginary time, whereas $t$ controls the perturbation strength. The derivative obeys
$\partial_uT_t=T_tA_t(u)$ with
$A_t(u)\coloneqq-(t\beta g/2)e^{-u\beta H_0/2}Ve^{u\beta H_0/2}$.
Because $P_\ell(a)$ commutes with $H_0$, the expansion of
$T_t(1)P_\ell(a)T_t(1)^{-1}$ is a connected nested-commutator expansion for $Q_\ell(t,a)$.

After $i$ insertions at most $1+ib_h$ physical sites are reached, so at most $C(1+ib_h)$ enlarged supports of subsequent insertions can overlap the reached set. At order $j$, the ordered-time simplex supplies $1/j!$, canceling this factorial growth. The same support and norm bounds hold for both signs of imaginary time. Let $B(\ell,r)$ denote the physical cells within lattice distance $r$ of $\ell$. Let $r_*$ be a fixed upper bound on the distance added by one overlapping enlarged support. This yields
\begin{equation}
 Q_\ell(t,a)=P_\ell(a)+\sum_{j\ge1}t^jQ_{\ell,j}(a),
 \label{sm:eq:parent-series}
\end{equation}
with
\begin{equation}
 \norm{Q_{\ell,j}(a)}\le\frac{\lambda^j}{j!}\prod_{i=0}^{j-1}C(1+ib_h)\le(C'\lambda)^j,
 \qquad \supp Q_{\ell,j}(a)\subset B(\ell,jr_*).
 \label{sm:eq:parent-series-bound}
\end{equation}
The constants are uniform in the product-basis label $a$.

Set $\lambda_{\rm aux}\coloneqq C\lambda$, absorbing the preceding constants. For fixed $a$, group terms in
$Q_\ell^\dagger Q_\ell-P_\ell$ by largest insertion order to write $\mathcal K_a(t)=\sum_\ell P_\ell(a)+\sum_{\ell,j}W_{\ell,j}(t)$. These Hermitian terms are supported on $B_{\ell,j}\coloneqq B(\ell,jr_*)$, with the label $a$ suppressed. For $\lambda_{\rm aux}\le1/4$, they obey
\begin{equation}
 \norm{W_{\ell,j}(t)}\le C\lambda_{\rm aux}^j,
 \qquad \norm{\partial_tW_{\ell,j}(t)}\le Cj\lambda_{\rm aux}^j.
 \label{sm:eq:parent-shells}
\end{equation}
Fix $\mu>0$ and reduce the small-coupling threshold if necessary. The number of roots whose radius-$jr_*$ block contains a site, and the size of each block, are both $O((1+j)^d)$. Consequently
\begin{align}
 &\sup_x\sum_{\ell,j:x\in B_{\ell,j}}|B_{\ell,j}|e^{\mu\diam B_{\ell,j}}
   (\norm{W_{\ell,j}}+\norm{\partial_tW_{\ell,j}})
 \nonumber\\
 &\qquad\le C\sum_{j\ge1}(1+j)^{2d+1}
             e^{2\mu jr_*}\lambda_{\rm aux}^j=O(\lambda_{\rm aux}).
 \label{sm:eq:parent-local-norm}
\end{align}

To obtain a uniform gap, apply the product-state case of the commuting-projector gap-stability theorem of Ref.~\cite{SM-BravyiHastings2011}. Its reference Hamiltonian is the product parent $\sum_\ell P_\ell(a)$ with a one-dimensional ground space. Every local reference gap is one. Its local ground spaces consist of the same product labels, giving the required local consistency and local indistinguishability conditions. The physical cell sites can be embedded in a fixed refinement of the cubic lattice; enclosing the shells in cubes gives the exponentially decaying interaction decomposition controlled by Eq.~\eqref{sm:eq:parent-local-norm}. Onsite basis changes relate different $a$ without changing these estimates. The theorem and the exact one-dimensional kernel imply
\begin{equation}
 \operatorname{spec}\mathcal K_a(t)
 \subset\{0\}\cup[\Delta,\infty),\qquad\Delta>0,
 \label{sm:eq:aux-gap}
\end{equation}
uniformly in $L,a,t$. All smallness conditions can be met by a fixed upper bound on $\lambda$.

\subsection{Controlled spectral flow and a local circuit}
\label{sm:app:controlled-flow}

The endpoint spectral flow will implement the conditional map
\begin{equation}
 \ket a_{\rm C}\ket a_{\rm P}
 \longmapsto e^{i\theta_a}\ket a_{\rm C}\ket{\psi_{1,a}}_{\rm P}.
 \label{sm:eq:controlled-flow}
\end{equation}
Here C denotes the control registers and P the physical spins; the phase disappears when the controls are discarded.

Attach one binary control to each physical site and define the mismatch projector
\begin{equation}
 \widetilde P_\ell\coloneqq I-
 \sum_{r=0}^1\proj r_{{\rm C},\ell}\otimes\proj r_{{\rm P},\ell}.
 \label{sm:eq:control-projector}
\end{equation}
Conjugate this local projector on the physical factors and form the positive sum
\begin{align}
 \widetilde Q_\ell(t)&\coloneqq
 (I_{\rm C}\otimes M_t)\widetilde P_\ell
 (I_{\rm C}\otimes M_t^{-1}),\label{sm:eq:controlled-local-term}\\
 \widetilde{\mathcal K}(t)&\coloneqq
 \sum_\ell\widetilde Q_\ell(t)^\dagger\widetilde Q_\ell(t)
 =\sum_a\proj a_{\rm C}\otimes\mathcal K_a(t).
 \label{sm:eq:controlled-parent}
\end{align}
In each fixed control sector, $\widetilde P_\ell$ reduces to $P_\ell(a)$, proving the last equality. The first expression allows locality to be proved term by term; the second identifies the spectrum sector by sector.
The same commutator argument proves its local interaction estimate directly: no sum of norms over exponentially many labels is taken. Equation~\eqref{sm:eq:aux-gap} supplies the gap above its zero-energy band by direct sum. The product-gap theorem was applied to each parent with a unique ground state, not to the locally distinguishable controlled band.

Quantitative spectral flow~\cite{SM-BachmannEtAl2012} along this uniformly gapped path gives unitaries $U(t)$ mapping the initial ground space to that at $t$. We use dimensionless path time $t\in[0,1]$ and the convention
\begin{align}
 D(t)&\coloneqq\sum_XD_X(t),\qquad \supp D_X(t)\subseteq X,
 \label{sm:eq:flow-generator}\\
 \frac{dU(t)}{dt}&=iD(t)U(t),\qquad U(0)=I.
 \label{sm:eq:flow-convention}
\end{align}
Here $X$ is a set of paired physical and control sites, each $D_X(t)$ is Hermitian, and $U\coloneqq U(1)$ is the endpoint unitary. The spectral-flow generator $D(t)$ is dimensionless. Both $\widetilde{\mathcal K}(t)$ and its derivative commute with every control projector. The filtered time evolutions used to construct $D(t)$ preserve this commutation, so $D(t)$ also commutes with those projectors. Hence the endpoint unitary $U$ acts on the unique ground state for each control label as in Eq.~\eqref{sm:eq:controlled-flow}, up to a phase that disappears after discarding the control.

To approximate this flow by a finite circuit, use the generator's almost-exponential locality in the form
\begin{equation}
 \sup_x\sum_{X\ni x:\diam X>r}\norm{D_X(t)}
 \le Ce^{-c\sqrt r}.
 \label{sm:eq:flow-tail}
\end{equation}
Truncating the generator changes the endpoint unitary by at most $Cn e^{-c\sqrt r}$. Allocate error $\varepsilon/4$ to this truncation and choose sufficiently large constants in
\begin{equation}
 r\coloneqq C_1\log^2(n/\varepsilon),\qquad
 \ell_{\rm ov}\coloneqq C_2\log(nr^d/\varepsilon).
 \label{sm:eq:flow-cutoffs}
\end{equation}
Here $r$ is the physical range cutoff, and $\ell_{\rm ov}$ is an overlap width in coarse-block units. Coarse-grain into blocks of side $O(r)$. The truncated interactions have fixed range in block units and strength $O(r^d)$ per block. Divide the unit flow interval into $O(r^d)$ slices so each slice has bounded block strength. A Lieb--Robinson block decomposition with overlap $\ell_{\rm ov}$~\cite{SM-HaahEtAl2021} then approximates every slice, allocating its errors so their sum is at most $\varepsilon/4$. Together with the truncation error, this gives a total operator-norm error at most $\varepsilon/2$. For fixed $d$, each slice requires a bounded number of layers, with gate diameter $O(\ell_{\rm ov})$ in block units. The resulting unitary circuit obeys
\begin{align}
 \norm{U-U_{\rm cir}}&\le\varepsilon/2,
 \label{sm:eq:flow-operator-error}\\
 \text{depth}&=O(r^d),\\
 \text{gate diameter}&=O(r\ell_{\rm ov}),\\
 R_{U_{\rm cir}}&=O(r^{d+1}\ell_{\rm ov}).
 \label{sm:eq:flow-circuit-size}
\end{align}
Only arbitrary within-block gates are asserted. The time slicing is needed because the coarse-grained block strength grows with $r$.

Copy the diagonal input $\tau_g$ into product-initialized controls using onsite controlled-NOT gates, apply $U_{\rm cir}$, and trace out the controls. For any density matrix $\omega$,
\begin{equation}
 \norm{U_{\rm cir}\omega U_{\rm cir}^\dagger-U\omega U^\dagger}_1
 \le2\norm{U_{\rm cir}-U}\le\varepsilon.
 \label{sm:eq:unitary-trace-error}
\end{equation}
This follows by adding and subtracting $U\omega U_{\rm cir}^\dagger$ and using unitary invariance of trace norm. Partial trace cannot increase this error. Together with Eq.~\eqref{sm:eq:conditional-ensemble}, it gives a deterministic channel $\mathcal R_{\varepsilon}^{\rm coh}$ satisfying
\begin{equation}
 \norm{\mathcal R_{\varepsilon}^{\rm coh}(\tau_g)
             -\rho_{\beta,g,L}}_1\le\varepsilon.
 \label{sm:eq:recovery-error}
\end{equation}
The correlated label distribution comes from the input, while the auxiliary registers were initially independent. Combining Eq.~\eqref{sm:eq:recovery-error} with the classical encoder and combining dephasing with the classical decoder proves the quantum part of Theorem~\ref{sm:thm:main}. Taking $\varepsilon=L^{-s}$ gives the claimed global accuracy and polylogarithmic causal radius.
\Needspace{9\baselineskip}
\section{High-temperature preparation and independent physical ancillas}
\label{sm:app:high-ancilla}

The auxiliary-parent method also gives a direct high-temperature preparation without using gauge structure. We first prove that statement, then check the changes caused by independent onsite systems in the physical Hamiltonian.

The companion Letter's high-temperature result~\cite{SM-WatanabeThermodynamics2026} controls free-energy analyticity by a normalized-trace expansion. The preparation result here instead requires an auxiliary gapped path and a local circuit with vanishing whole-system error; these are constructed below using the methods of Sec.~\ref{sm:app:quantum}.

\begin{theorem}[High-temperature local preparation]
\label{sm:thm:high}
For any fixed $d,m$ with $0\le m\le d-1$ and bounded finite-range interaction class above, there is $\tilde a'>0$ such that
\begin{equation}
 \beta[2(d-m)J+n_{\mathrm{inc}}|g|]<\tilde a'
 \label{sm:eq:high-domain}
\end{equation}
implies
\begin{equation}
 \boldsymbol\rho_{\beta,g}\sim_{\LC}\boldsymbol\rho_\infty.
 \label{sm:eq:high-equivalence}
\end{equation}
Both directions obey the same locality, product-ancilla, and whole-system accuracy guarantees as in Theorem~\ref{sm:thm:main}.
\end{theorem}
The purification uses one onsite work qubit per physical qubit and the product Bell reference
\begin{equation}
 \ket\Omega\coloneqq\bigotimes_\ell\ket{\Omega_\ell},\qquad
 \ket{\Omega_\ell}\coloneqq\frac{\ket{00}_\ell+\ket{11}_\ell}{\sqrt2},
 \label{sm:eq:bell-reference}
\end{equation}
where the first entry is physical and the second is its work partner.

\subsection{A product purification and its auxiliary gap}

Write $H_g=\sum_Yh_Y$, with support size at most $b'\coloneqq\max\{2(m+1),b\}$ and uniformly bounded range. Define the local energy scale
\begin{equation}
 j_g\coloneqq2(d-m)J+n_{\mathrm{inc}}|g|,
 \qquad \lambda_{\rm h}\coloneqq\beta b'j_g.
 \label{sm:eq:app-local-energy}
\end{equation}
Then $\sup_\ell\sum_{Y\ni\ell}\norm{h_Y}\le j_g$.

Use the Bell factors of Eq.~\eqref{sm:eq:bell-reference}, with P denoting the physical qubit and W its onsite work partner. Each pair is a four-dimensional local site. The thermal-parent construction~\cite{SM-FeiguinKlich2013} gives
\begin{align}
 M_t^{\rm h}&\coloneqq e^{-t\beta H_g/2}\otimes I_{\rm W},
 \label{sm:eq:high-M}\\
 \ket{\Xi_t}&\coloneqq
 \frac{M_t^{\rm h}\ket\Omega}
      {\sqrt{\Tr e^{-t\beta H_g}/2^n}},
 \label{sm:eq:high-purification}\\
 P_\ell^{\rm h}&\coloneqq I-\proj{\Omega_\ell},
 \label{sm:eq:high-P}\\
 \mathcal K^{\rm h}(t)&\coloneqq
 \sum_\ell(M_t^{\rm h}P_\ell^{\rm h}(M_t^{\rm h})^{-1})^\dagger
       (M_t^{\rm h}P_\ell^{\rm h}(M_t^{\rm h})^{-1}).
 \label{sm:eq:high-parent}
\end{align}
Tracing out work qubits from $\proj{\Xi_t}$ gives the Gibbs state at inverse temperature $t\beta$. The parent kernel is exactly the span of $\ket{\Xi_t}$, and the reference parent $\sum_\ell P_\ell^{\rm h}$ has gap one on a product of four-dimensional paired sites.

To bound the parent's range and strength, define for operators $H,O$
$\operatorname{ad}_H^0(O)\coloneqq O$ and
$\operatorname{ad}_H^{j+1}(O)\coloneqq[H,\operatorname{ad}_H^j(O)]$ for $j\ge0$, with $[H,O]\coloneqq HO-OH$.
The nested-commutator expansion satisfies
\begin{equation}
 \left\|\frac{(-\beta/2)^j}{j!}
       \operatorname{ad}_{H_g\otimes I}^j(P_\ell^{\rm h})\right\|
 \le(\beta b'j_g)^j.
 \label{sm:eq:high-commutator-bound}
\end{equation}
After $i$ commutators at most $1+ib'$ paired sites are reached. The sum of norms of possible next terms is at most $j_g(1+ib')$, and every commutator contributes a factor two. Multiplying these bounds and dividing by $j!$ proves Eq.~\eqref{sm:eq:high-commutator-bound}. The supports lie in radius $O(j)$.

Grouping the parent into shells therefore proves Eq.~\eqref{sm:eq:parent-local-norm} with $\lambda_{\rm aux}$ replaced by a constant times $\lambda_{\rm h}$. The same product-gap theorem, spectral flow, and block circuit prepare $\ket{\Xi_1}$ with any fixed inverse-power trace accuracy when $\beta j_g<\tilde a'$, absorbing fixed $b'$ into the constant. All Bell pairs are prepared onsite from product inputs. Discarding work sites prepares $\rho_{\beta,g,L}$, while onsite depolarization supplies the exact reverse channel. This proves Theorem~\ref{sm:thm:high}.

For the unperturbed classical model, the single-site heat-bath influence row sum obeys
\begin{equation}
 \alpha\le2(d-m)(2m+1)\tanh(\beta J).
 \label{sm:eq:dobrushin-unperturbed}
\end{equation}
When this bound is below one, $t$ complete colored sweeps give whole-system trace error at most $2n\alpha^t$, so $t=O(\log L)$ suffices for any fixed inverse-power accuracy.

For the explicit classical bound in Eq.~\eqref{sm:eq:dobrushin-unperturbed}, changing one neighboring spin in an interaction changes the single-spin conditional field by $2\beta J$. The associated total variation influence is at most $\tanh(\beta J)$. For $L\ge3$, two distinct interaction cells containing $\ell$ share no other boundary $m$-cell, so these neighbors are distinct. Summing over $2(d-m)$ interaction cells and $2m+1$ neighbors per interaction gives that row bound. A finite coloring makes simultaneous updates independent within each color. Coupling a sweep to stationarity contracts the maximum disagreement probability by $\alpha$, so the probability of any disagreement is at most $n\alpha^t$. Diagonal trace norm is twice total variation, giving $2n\alpha^t$ as stated.

\subsection{Physical ancillary assumptions and statement}
Disposable work registers should be distinguished from extra physical degrees of freedom whose joint Gibbs state is part of the target. Suppose a fixed finite number of bounded-dimensional ancillary sites per lattice cell have independent onsite Hamiltonians,
\begin{equation}
 H_A\coloneqq\sum_jh_j,\qquad
 \tau_A\coloneqq\bigotimes_j\frac{e^{-\beta h_j}}{\Tr e^{-\beta h_j}}.
 \label{sm:eq:physical-ancillas}
\end{equation}
Let $\Delta_A$ bound their spectral widths. The joint Hamiltonian is $H_0+H_A+gV$, with identity factors understood. The terms in $gV$ may include system--ancilla and ancilla--ancilla couplings, with the same fixed locality conventions on the joint sites. The low-temperature reference becomes $\rho_{\rm flat,L}\otimes\tau_A$. For the classical branch, fix an onsite eigenbasis of each $h_j$ and require $V$ to be diagonal in its tensor product with the system $Z$ basis.

\begin{theorem}[Stability with physical ancillary systems]
\label{sm:thm:ancilla}
For $1\le m\le d-2$ and the joint model just specified, there are positive class-dependent constants $\tilde B_A,\tilde a_A,\tilde C_A,\tilde a'_A$ such that the sufficient low-temperature windows
\begin{equation}
 \beta J>\tilde B_A,
 \label{sm:eq:ancilla-temperature}
\end{equation}
\begin{equation}
 \begin{cases}
 \beta|g|<\tilde a_A,&\text{classical},\\
 \beta|g|e^{\tilde C_A\beta(J+\Delta_A)}<\tilde a_A,
                  &\text{Hermitian}.
 \end{cases}
 \label{sm:eq:ancilla-window}
\end{equation}
imply two-way conversion of the full joint state to $\boldsymbol\rho_{\rm flat}\otimes\boldsymbol\tau_A$. At high temperature,
\begin{equation}
 \beta[2(d-m)J+\Delta_A/2+n_{\mathrm{inc}}|g|]<\tilde a'_A
 \label{sm:eq:ancilla-high-window}
\end{equation}
implies two-way equivalence to an onsite product state, with fixed site multiplicities absorbed into the constants. All conversions have product work-register inputs, polylogarithmic causal radius, and whole-system trace-norm error $O(L^{-s})$ for each fixed $s>0$.
\end{theorem}
No gap or nondegeneracy of $h_j$ is required. Interacting ancillary reference states are outside this extension.

\subsection{Diagonal physical couplings}

Let $z$ denote the ancillary labels in the chosen onsite eigenbases. Their reference distribution $\tau_A(z)$ is a product. For a diagonal perturbation of the joint Hamiltonian $H_0+H_A+gV$, put $F(a,z)\coloneqq-\beta gV(a,z)$. The joint law is proportional to
\begin{equation}
 e^{-\beta H_0(a)}\tau_A(z)e^{F(a,z)}.
 \label{sm:eq:ancilla-joint-law}
\end{equation}
The system marginal contains the exact exponent
\begin{equation}
 F_{\rm eff}(a)\coloneqq
      \log\mathbb E_{z\sim\tau_A}e^{F(a,z)}.
 \label{sm:eq:ancilla-effective}
\end{equation}
It is essential to take this logarithm, rather than the average of $F$.
This product-expectation reduction and its weighted connected expansion are also used for physical ancillas in Ref.~\cite{SM-WatanabeThermodynamics2026}. To convert the full joint state, we additionally need the conditional ancillary sampler constructed below.

To see why the expansion survives the ancillary sum, first retain finitely many generating labels and write $F=\sum_\alpha f_\alpha$, where $f_\alpha$ is the dimensionless local exponent on its assigned carrier. For the original perturbation these are $-\beta g v_X(a,z)$. For weighted interactions, require the joint-site analogue of Eq.~\eqref{sm:eq:weighted-norm}, $\sup_x\sum_{\alpha:x\in K_\alpha}\|f_\alpha\|_\infty e^{S_{\rm in}\nu_\alpha}$, to be small; $x$ ranges over system and ancillary sites, with the same linear carrier bounds. Choose the fixed input reserve $S_{\rm in}$ larger than the output reserve $S$ by enough to absorb the tree and sign-character expansion factors below.

Expand each $e^{f_\alpha}$ as $1+(e^{f_\alpha}-1)$ and join selected labels whose carriers overlap. Product expectation over $z$ factors between distinct connected groups. A connected label set $\mathcal A$ therefore has the activity given by the expectation below, with the bound
\begin{equation}
 \left\|\mathbb E_{\tau_A}\prod_{\alpha\in\mathcal A}
                    (e^{f_\alpha}-1)\right\|_\infty
 \le\prod_{\alpha\in\mathcal A}(e^{\norm{f_\alpha}_\infty}-1).
 \label{sm:eq:ancilla-activity-bound}
\end{equation}
The expansion is an exact hard-core gas of these groups. Its connected logarithm obeys the rooted-tree and local bounds of Sec.~\ref{sm:app:flux-potential}. Expanding its remaining system dependence in sign characters costs an exponential in the generating cost, absorbed by the reserve as in Eq.~\eqref{sm:eq:quantum-Fourier}. Thus the resulting $Q_S$ is $O(\beta|g|)$ for the original local perturbation. The same argument applies to weighted interactions, uniformly in the finite-label truncation, so their absolutely convergent limit is allowed. Carriers retain the integrated ancillary sites even when the remaining system support is smaller; this keeps the locality estimate valid. Applying Proposition~\ref{sm:prop:weighted} to the system marginal yields its conversion to the flat reference.

Restoring the full joint state requires the ancillary conditional law
\begin{equation}
 Q_a^A(z)\coloneqq
 \frac{\tau_A(z)e^{F(a,z)}}{\mathbb E_{\tau_A}e^{F(a,z)}}.
 \label{sm:eq:ancilla-conditional}
\end{equation}
The required influence bound is independent of small onsite probabilities. For any fixed probability weight $\nu$, let $q_f$ and $q_h$ be the normalized laws proportional to $\nu e^f$ and $\nu e^h$. Define $\operatorname{osc}u\coloneqq\sup u-\inf u$. Interpolate using the normalized law $q_\theta\propto\nu e^{h+\theta(f-h)}$, $0\le\theta\le1$. Differentiation gives $\partial_\theta q_\theta(z)=q_\theta(z)[(f-h)(z)-\mathbb E_{q_\theta}(f-h)]$. Taking half its $\ell^1$ norm and integrating proves
\begin{equation}
 \norm{q_f-q_h}_{\rm TV}\le\tfrac12\operatorname{osc}(f-h).
 \label{sm:eq:ancilla-tilt-bound}
\end{equation}
Changing one neighboring ancillary label alters only interaction terms meeting both sites. Each such term contributes $O(\beta|g|)$ to this oscillation. The fixed support and incidence bounds therefore give an influence row sum $O(\beta|g|)$, uniformly in $a$ and in the reference probabilities. For weighted interactions the row is bounded by a constant times $\sum_{\alpha:x\in K_\alpha}|K_\alpha|\|f_\alpha\|_\infty$. The factor $|K_\alpha|\le b_w\nu_\alpha$ is absorbed by the exponential reserve in the joint-site bound above.

Cutoffs and colored heat-bath sweeps give a conditional sampler with the same polylogarithmic locality and inverse-power whole-state accuracy as in Sec.~\ref{sm:app:orbit}. The forward channel discards ancillary labels, transports the system marginal, and appends $\tau_A$. The reverse channel transports the system part and samples Eq.~\eqref{sm:eq:ancilla-conditional}. It returns the full joint distribution.

\subsection{Quantum physical couplings and high temperature}

For general Hermitian couplings, replace the commuting reference $H_0$ by $H_0+H_A$ and use its onsite product eigenbasis. A perturbation term meets only a fixed number of ancillary sites. After centering each onsite spectrum, conjugation by those terms contributes at most $e^{C\beta\Delta_A}$, while conjugation by $H_0$ is bounded as in Eq.~\eqref{sm:eq:lambda}. The connected diagonal expansion consequently has strength at most
\begin{equation}
 C\beta|g|e^{C'\beta(J+\Delta_A)}.
 \label{sm:eq:ancilla-quantum-strength}
\end{equation}

The resulting diagonal law includes weak exponentially decaying system--ancilla interactions; the preceding marginal and conditional expansions apply with their generating carriers retained. Auxiliary conditional vectors and parents with a one-dimensional ground space are defined in the joint onsite basis. The reference parents still have gap one, independent of the physical onsite gaps. The commutator, gap-stability, controlled-flow, and trace-error arguments of Sec.~\ref{sm:app:quantum} apply with fixed larger onsite dimension. This proves Eq.~\eqref{sm:eq:ancilla-window} for the full joint density matrix.

At high temperature, center each onsite $h_j$ so its operator norm is at most $\Delta_A/2$. The per-site interaction bound in the purification proof is then bounded by a class-dependent multiple of
$2(d-m)J+\Delta_A/2+n_{\mathrm{inc}}|g|$. Fixed numbers of physical ancillary sites per cell and support-size constants are absorbed into $\tilde a'_A$. The same construction proves the high-temperature part of Theorem~\ref{sm:thm:ancilla}. Degenerate onsite eigenvalues cause no difficulty. Independent onsite reference terms are essential here; no gap or mixing theorem for an arbitrary interacting ancillary system is being assumed.
\section{Reader continuity, locality, and the canonical flat bit}
\label{sm:app:invariant}

This section works in the stability range $1\le m\le d-2$ and checks the optimization over all readers, including channels with discarded registers. It also explains why the canonical all-sector ensemble is part of the reference specification.

\Needspace{9\baselineskip}
\subsection{Reader geometry and invariant statement}
Fix a coordinate orientation, say the first $m$ axes. Let $C_{1,L}$ and $C_{2,L}$ be two parallel primitive coordinate $m$-tori, each wrapping once around those axes, and separated transversely by at least $\kappa L$ for a fixed $\kappa>0$. For a dimensionless relative width $\zeta>0$, the reader region $R_{i,L}(\zeta)$ contains the physical $m$-cells within transverse distance $\zeta L$ of $C_{i,L}$, including cells intersecting the edge of this neighborhood. This last convention adds only a bounded number of lattice spacings. For sufficiently small fixed $\zeta$, the regions are disjoint, periodic in the first $m$ directions, and bounded in the transverse directions. The orientation, positions, and separation convention are fixed independently of the state.

For binary Hermitian readers $A_i$, define
\begin{align}
 F_{\zeta,L}^{(m)}(\rho)&\coloneqq
 \sup_{\substack{A_i=A_i^\dagger,\ A_i^2=I\\
             \supp A_i\subset R_{i,L}(\zeta)}}
 \left|\Tr\rho A_1A_2-\Tr\rho A_1\Tr\rho A_2\right|,\label{sm:eq:finite-reader}\\
 \mathcal F_m(\boldsymbol\rho)&\coloneqq
 \lim_{\zeta\downarrow0}\limsup_{L\to\infty}
                  F_{\zeta,L}^{(m)}(\rho_L).
 \label{sm:eq:F-definition}
\end{align}
The thermodynamic limit is taken before the relative width tends to zero. The outer limit exists by monotonicity in the width, and Cauchy--Schwarz gives $0\le\mathcal F_m\le1$.
\begin{proposition}[Monotonicity and two-way invariance]
\label{sm:prop:F-monotone}
If $\boldsymbol\rho\to_{\LC}\boldsymbol\sigma$, then
$\mathcal F_m(\boldsymbol\sigma)\le\mathcal F_m(\boldsymbol\rho)$.
Consequently $\mathcal F_m$ is constant on every two-way local-channel equivalence class.
\end{proposition}
For a closed physical $m$-cell set $C$, write $W_C\coloneqq\prod_{\ell\in C}Z_\ell$. The flat reference values established below are
\begin{equation}
 \langle W_{C_1}W_{C_2}\rangle_{\rm flat}=1,
 \qquad \langle W_{C_i}\rangle_{\rm flat}=0,
 \qquad \mathcal F_m(\boldsymbol\rho_{\rm flat})=1.
 \label{sm:eq:F-flat}
\end{equation}
Product inputs instead have $\mathcal F_m=0$.

\subsection{Continuity and pullback through a local channel}

For two normalized states with $\varepsilon\coloneqq\norm{\rho-\sigma}_1$, any norm-one observable has an expectation difference at most $\varepsilon$. If $A_1,A_2$ are Hermitian contractions on disjoint regions, their pair expectation changes by at most $\varepsilon$, and their product of means changes by at most $2\varepsilon$. Thus their covariances differ by at most $3\varepsilon$. Supremizing over the same readers preserves this continuity bound.

Optimizing over Hermitian contractions is equivalent to optimizing over the binary readers in Eq.~\eqref{sm:eq:finite-reader}. Any finite-dimensional Hermitian contraction $A$ is a convex combination of Hermitian involutions: in its eigenbasis, independently average signs whose means equal its eigenvalues. Covariance is separately linear in each observable, and its absolute value is bounded by the largest absolute covariance of a pair of extremal choices.

Now take a local circuit $\Phi_L$ implementing a conversion. For fixed $\zeta>0$, its sublinear causal radius makes the full backward cones of the two output readers disjoint for all sufficiently large $L$. Use a local dilation of every channel gate, including every classical message and work register. Pull the two observables backward through that dilation. They have disjoint input supports, so their joint pullback is their product before averaging over work inputs. The work registers start in a product state over local sites. Averaging over the two disjoint work sets therefore yields two Hermitian contractions $B_1,B_2$ on the corresponding physical backward cones, with
\begin{align}
 \Tr\Phi_L(\rho_L)A_1A_2&=\Tr\rho_LB_1B_2,\\
 \Tr\Phi_L(\rho_L)A_i&=\Tr\rho_LB_i.
 \label{sm:eq:reader-pullback}
\end{align}
This restricted factorization follows from the disjoint dilation. In general $\Phi_L^*(A_1A_2)$ is not equal to $\Phi_L^*(A_1)\Phi_L^*(A_2)$ without that geometric condition.

The backward cones fit in $R_{i,L}(2\zeta)$ once $L$ is large. Write $\varepsilon_L\coloneqq\norm{\Phi_L(\rho_L)-\sigma_L}_1$ for the conversion error. Continuity gives
\begin{equation}
 F_{\zeta,L}^{(m)}(\sigma_L)
 \le F_{2\zeta,L}^{(m)}(\rho_L)+3\varepsilon_L.
 \label{sm:eq:reader-monotone-finite}
\end{equation}
Taking $\limsup_{L\to\infty}$ first and then $\zeta\downarrow0$ proves Proposition~\ref{sm:prop:F-monotone}. The same argument absorbs a preliminary recovery channel into the reader-width budget. Product physical inputs factor over the two cones, proving the zero value for all optimized readers, rather than only for Wilson surfaces.

\subsection{Flat signs, local indistinguishability, and fixed sectors}

Two parallel primitive $m$-tori bound a cellular slab; multiplying the flat constraints across it proves $W_{C_1}=W_{C_2}$. To flip this common sign, choose the physical $m$-cells pierced by a complementary dual coordinate $(d-m)$-torus. The dual sheet has no boundary, so flipping its pierced physical cells leaves every $B_p$ unchanged. It meets each reader torus with odd parity and hence reverses both Wilson signs. This bijection of the full flat set establishes their zero individual means and Eq.~\eqref{sm:eq:F-flat}.

The same construction explains local indistinguishability. The location of a complementary sheet may be shifted away from any fixed bounded contractible region without changing the corresponding winding flip. Hence pairing the flat configurations between two winding sectors can be done without changing the configuration in that region. All such sector marginals are equal there. A local gauge flip on the cells containing one $(m-1)$-cell cannot change a closed Wilson surface: its intersection with that surface has even parity.

For clarity, the full set of flat configurations can be generated by choosing one of the $2^{\binom dm}$ coordinate winding-sign assignments, a deterministic flat representative for it, and independent uniform gauge signs $\eta_u$ on all $(m-1)$-cells $u$, acting as in Eq.~\eqref{sm:eq:gauge-action}. The gauge action reaches every flat configuration with the specified signs.

To verify completeness, compare two flat configurations with the same Wilson winding signs. Their difference is flat and has all winding signs $+1$. Sweep this difference through a fundamental box, using the local construction in Sec.~\ref{sm:app:fillings}. Flatness cancels interior obstructions; the only possible residual signs occur on coordinate wrapping surfaces. Since these signs are all $+1$, the residue vanishes and local gauge flips generate the difference. Redundancies among the gauge parameters have equal multiplicity, since any two solutions differ by a parameter acting trivially. Uniform parameters consequently give the uniform ensemble in that sector.

A fixed-sector flat ensemble can thus be prepared from a deterministic representative by a bounded-depth circuit of independent gauge flips. It has $\mathcal F_m=0$ under the present product-input convention. The nonzero reference value comes from the canonical mixture over winding sectors, whose shared random bit cannot be supplied by local independent coins. Equality of $\mathcal F_m$ is a necessary condition for two-way equivalence, not a general completeness theorem for this invariant.

% ===== Supplemental bibliography =====
\begingroup
\makeatletter
\let\@FMN@list\@empty
\let\combined@label\label
\def\label#1{\combined@label{SM-bib-#1}}
\makeatother

\endgroup
\end{document}